\documentclass[12pt, a4paper, onecolumn]{article}
\usepackage[utf8]{inputenc}
\usepackage{amsmath}
\usepackage{graphicx} 
\usepackage{pdfpages}

\title{Emission and Coherent Control of Levitons in Graphene}
\author{
    A. Assouline\textsuperscript{1,*}, L. Pugliese\textsuperscript{1,2,*}, H. Chakraborti\textsuperscript{1,*}, \\
    Seunghun Lee\textsuperscript{3}, L. Bernabeu\textsuperscript{1,2}, M. Jo\textsuperscript{1}, K. Watanabe\textsuperscript{4}, \\
    T. Taniguchi\textsuperscript{4}, D. C. Glattli\textsuperscript{1}, N. Kumada\textsuperscript{5}, \\
    H.-S. Sim\textsuperscript{3}, F. D. Parmentier\textsuperscript{1}, P. Roulleau\textsuperscript{1,\textdagger}
}
\date{}

\begin{document}

\maketitle

\noindent\textsuperscript{1}SPEC, CEA, CNRS, Université Paris-Saclay, CEA Saclay, 91191 Gif-sur-Yvette Cedex, France. \\
\noindent\textsuperscript{2}Université Paris-Saclay, CNRS, Centrale Supélec, 91191 Gif-sur-Yvette Cedex, France. \\
\noindent\textsuperscript{3}Department of Physics, Korea Advanced Institute of Science and Technology, Daejeon 34141, Korea. \\
\noindent\textsuperscript{4}National Institute for Materials Science, 1-1 Namiki, Tsukuba 305-0044, Japan. \\
\noindent\textsuperscript{5}NTT Basic Research Laboratories, NTT Corporation, 3-1 Morinosato-Wakamiya, Atsugi 243-0198, Japan. \\
\noindent\textsuperscript{*}These authors contributed equally to this work. \\
\noindent\textsuperscript{\textdagger}Corresponding author. Email: preden.roulleau@cea.fr \\

\textbf{Abstract}
Flying qubits encode quantum information in propagating modes instead of stationary discrete states. Although photonic flying qubits are available, the weak interaction between photons limits the efficiency of conditional quantum gates. Conversely, electronic flying qubits can use Coulomb interactions, but the weaker quantum coherence in conventional semiconductors has hindered their realization. In this work, we engineered on-demand injection of a single electronic flying qubit state and its manipulation over the Bloch sphere. The flying qubit is a Leviton propagating in quantum Hall edge channels of a high-mobility graphene monolayer. Although single-shot qubit readout and two-qubit operations are still needed for a viable manipulation of flying qubits, the coherent manipulation of an itinerant electronic state at the single-electron level presents a highly promising alternative to conventional qubits.\\

Flying qubit experiments rely on the ability to encode information into a propagating state of a single photon or electron excitation, manipulate the information, and read it after operations~\cite{bauerle2018}. Photon flying qubits evolve in the vacuum, which substantially reduces decoherence processes. Conversely, electron flying qubits naturally experience a strong and tunable Coulomb interaction, which leads to easier two-qubit operations but gives rise to finite decoherence. 

Electronic flying qubits can benefit from recent breakthroughs of electron quantum optics in GaAs heterostructures, including the Mach-Zehnder interferometer (MZI)~\cite{ji2003, roulleau2008, yamamoto2012}, Hong-Ou-Mandel experiments~\cite{bocquillon2013, dubois2013}, robust high-fidelity single-electron sources based on Levitons (voltage pulses with a Lorentzian profile enabling pure-electron excitation without any hole)~\cite{dubois2013, glattli2017}, and the demonstration of single-electron quantum tomography~\cite{jullien2014}. Nevertheless, a basic quantum manipulation of an on-demand and propagating single electron is still missing~\cite{dasenbrook2015}, primarily owing to the short coherence length\cite{roulleau2008} of excited electrons in conventional semiconductors. This problem could be solved by using graphene, a two-dimensional atomically thin material, which shows outstanding coherence properties under relatively large bias~\cite{jo2021, jo2022}. Owing to its valley pseudospin degrees of freedom, graphene provides a very promising platform.

Recently, it was shown that the valley degrees of freedom in graphene can be addressed electrostatically~\cite{schaibley2016, mak2014, shimazaki2015, ju2015, li2018, gorbachev2014}. In particular, coherent and tunable electronic beam splitters, which couple quantum Hall edge channels with opposite valley polarizations, were formed~\cite{jo2021, morikawa2015, wei2017}. Then, an electronic MZI along a PN junction was realized by placing two valley beam splitters at both ends of the junction. However, quantum manipulation at the single-electron level, a crucial prerequisite for realizing electronic flying qubits, is still lacking in graphene. One primary reason is the absence of an electron pump, which generally requires dynamical control of quantum dots. This is extremely challenging in graphene because of the absence of a bandgap, unlike in conventional semiconductors.

In this work, we show that Levitons can be a reliable option for the injection of single electrons. Then, we demonstrate Bloch sphere rotation manipulations, taking advantage of the valley degrees of freedom in graphene MZI. This validates that the quantum coherence of the graphene MZI can be more than a few micrometers in length under high-frequency excitations. The information on final states is read statistically by combining conductance and noise measurements while periodically repeating the qubit operation.

\textbf{On-demand injection of single electrons in graphene}

In conventional semiconductors, different techniques for controlling on-demand electron excitation have been developed. The electron pump, composed of a series of tunnel barriers with an island or a dot, is one of them~\cite{kouwenhoven1991}. Fast manipulation of the tunnel barriers enables sequential emission of electrons. In this case, electrons are injected well above the Fermi energy, which should limit Coulomb interactions with electrons from the Fermi sea. However, because the electron excitations are far from the Fermi surface, there is more room to excite electron-hole pairs out of the ground state, leading to a very short relaxation length~\cite{rodriguez2020}. For graphene, although the development of well-defined quantum dots is an active field~\cite{eich2018}, their fabrication remains challenging and an electron pump has not been demonstrated. 

An alternative approach to on-demand single-electron injection is a direct application of a voltage pulse $V(t)$, where $t$ is time, on the emitter contact, with the condition that the Faraday flux $\phi(t) = e \int V(t') dt'/\hbar$ is an integer value, where $e$ is the charge of the electron and $\hbar$ is Planck’s constant. More specifically, it has been demonstrated that by shaping the pulse as a Lorentzian function, a single electron can be emitted without the creation of unwanted electron-hole pairs~\cite{ivanov1997,keeling2006}. This excitation has been called Leviton. In addition to its simplicity, this approach allows for the emission of electrons very close to the Fermi energy, where there is minimal room for electron-hole pair creation, thereby protecting the emitted electron from possible relaxation and decoherence. In the following sections, we present a demonstration of on-demand single-electron injection based on Levitons in graphene.

\textbf{Coherent manipulation and reading of the graphene qubit}

The next step was to demonstrate the coherent manipulation of single electrons during propagation. The most elementary quantum manipulation is the rotation of a single qubit on the Bloch sphere. This can be achieved through an electronic MZI, which can be formed in graphene by mixing two N- and P-type edge channels with opposite valley isospin $\pm\vec{\omega}$ in the bipolar quantum Hall regime~\cite{jo2021}. 

At the first electron beam splitter of the interferometer, the degree of valley-channel mixing can be characterized by a transmission probability $T_1$ and a reflection probability $R_1 = 1 - T_1$ of the beam splitter. The initial state of an electron defined after the first beam splitter can be written as a quantum-mechanical superposition 
\[
|\psi_{\text{initial}}\rangle = \cos\left(\frac{\theta_1}{2}\right)|\vec{\omega}\rangle + \sin\left(\frac{\theta_1}{2}\right)|-\vec{\omega}\rangle,
\]
where we introduce the channel mixing angle $\theta_1$ with $\cos\left(\frac{\theta_1}{2}\right) = r_1$ and $|r_1|^2 = 1 - T_1$, where $r$ is a reflection coefficient.

The valley superposition state evolves by acquiring the Aharonov-Bohm phase $\phi_{\text{AB}}$ (which is equal to $2\pi BA/\Phi_0$, where $B$ is the magnetic field, $A$ is the interferometer area, and $\Phi_0 = h/e$ is the flux quantum) along the MZI. The final state at arrival at the second beam splitter becomes 
\[
|\psi_{\text{final}}\rangle = \cos\left(\frac{\theta_1}{2}\right)|\vec{\omega}\rangle + \sin\left(\frac{\theta_1}{2}\right)e^{i\phi_{\text{AB}}}|-\vec{\omega}\rangle.
\]
Therefore, beam splitters combined with the Aharonov-Bohm effect enable the basic operations of a valley-isospin flying qubit.

After the electron passes through the second beam splitter, the final state is measured by projecting it on the output left state 
$\vert \tilde{\Psi} \rangle = \sin \left( \frac{\theta_2}{2} \right) \vert \omega \rangle + \cos \left( \frac{\theta_2}{2} \right) \vert - \omega \rangle$, 
with $\sin \left( \frac{\theta_2}{2} \right) = t_2$, 
where $T_2 = \vert t_2 \vert^2$ represents the transmission probability of the second beam splitter. 
We measured the transmission probability of the MZI as 
$T_{\text{MZI}} = \vert \langle \tilde{\Psi} \vert \psi_{\text{final}} \rangle \vert^2$ 
and the associated noise. 
Using the noise measurement, we show that the fundamental property of a Leviton, 
namely the minimization of the number of electron-hole pairs, is preserved during its propagation through the interferometer.

\textbf{Voltage-pulse generation}

The sample we used is depicted in Fig. 1A. 
A global graphite back gate and a metallic top gate (labeled as Top G in Fig. 1A) are deposited on the right half of the sample in order to independently tune the electron density in the left and right halves of the sample. 
An electronic MZI can be formed in the bipolar quantum Hall regime \cite{jo2021, morikawa2015, wei2017, handschin2017, makk2018, tworzydlo2007, trifunovic2019}. 
The filling factors of the N and P regions are set to $\nu_N = 2$ and $\nu_P = -2$, respectively, resulting in four copropagating channels along the PN junction. 
It has been shown that edge channels from the N and P regions with opposite valley polarization can be mixed by adjusting upper ($\text{SG}_1$) and lower ($\text{SG}_2$) side gates placed at the intersection between the PN junction and the physical edge of graphene \cite{jo2021}.

We first consider the case where the valley beam splitter is formed only at the upper edge, 
which can be obtained by setting the filling factors below the upper and lower side gates to $\nu_1 \leq -2$ and $\nu_2 = 0$, respectively. 
We constructed periodic Lorentzian pulses by summing a series of harmonics with controlled amplitude and phase. 
Because the amplitude and phase of the pulse change during the propagation along the electromagnetic lines in the cryostat, 
the pulse emitted at the output of the generator at room temperature and the one that propagates along the lines of the cryostat into the sample at base temperature are different (Fig. 1A). 
To resolve this issue, it was crucial to determine the amplitude and phase of each harmonic that is required to build periodic Lorentzian pulses “in situ” by measuring the photoassisted shot noise \cite{glattli2017}. 
To this end, we first considered the simplest case with a single mode ($\nu = 10.5~\mathrm{GHz}$). 
A sinusoidal potential at frequency $\nu$ is applied on the upper right ohmic contact ($C_R$), 
and the shot noise is measured at the lower left ohmic contact ($C_N$). 

By coherent scattering at the MZI, this generates a photoassisted shot noise, 
which is characterized by shot-noise singularities at $e V_{\text{dc}} = n h \nu$ \cite{reydellet2003, kapfer2019, gabelli2013}, 
where $V_{\text{dc}}$ is the superimposed dc bias, $n$ is an integer, and $\nu$ is frequency. 
The exact number of electron-hole pairs is computed by comparing measured photoassisted shot noise 
with established theoretical expectations \cite{dubois2013, deprez2021} 
[see also, “Theoretical predictions” section in \textbf{SI}].
The excess noise at the thermal equilibrium generated by the partitioning of electron-hole pairs at the first beam splitter can be expressed as 
\[
S_I = S_I^0 \sum_{l=-\infty}^{+\infty} \left[ \coth \left( \frac{l}{2\theta_e} \right) - 2\theta_e \right] J_l^2(\alpha),
\]
where $\theta_e = k_B T_e / h \nu$ is the temperature in frequency units, $\alpha = e V_{\text{ac}} / h \nu$ is the photon number, $J_l$ is the Bessel function ($l$ is an integer), and 
\[
S_I^0 = 2 \frac{2e^2}{h} D(1-D) h \nu
\] 
is the typical scale of the photoassisted shot noise ($D$ is the transmission probability of one edge channel). Note that the presence of the factor 2 is caused by the two modes involved in the partitioning at the $v_N = 2$ and $v_P = -2$ configurations. The agreement with the theoretical photoassisted shot noise $S_I$ confirms that the amplitude at the injection contact can be precisely determined (Fig. 1B). Next, to determine the phase, we measured the shot noise while varying the phase difference $\varphi$ between the two harmonics, which defines the biharmonic signal 
\[
V_{\text{ac}, bi}(q, t, \varphi) = q + \alpha_1 \cos(2\pi \nu t) + \alpha_2 \cos(2\pi \nu t + \varphi),
\]
where $q = e V_{\text{dc}} / h \nu$, $\alpha_1 = e V_{\text{ac}, 1} / h \nu$, and $\alpha_2 = e V_{\text{ac}, 2} / 2 h \nu$. In Fig. 1C, 
\[
\delta S_I = S_I(q = 1, \alpha_1, \alpha_2, \varphi) - S_I(q = -1, \alpha_1, \alpha_2, \varphi)
\]
is plotted as a function of $\varphi$ for $\nu = 3.5~\text{GHz}$. By comparing the results with the photoassisted theory in \textbf{SI}, we could accurately adjust $\varphi$ at the injection contact. 
We repeated this process for the third and fourth harmonics (see Fig. S10 for a $\nu - 3\nu$ calibration). Combined with the amplitude control, this result demonstrates that it is possible to engineer any pulse shape at the injection contact [see also the detailed discussion in section, ``Engineering voltage pulses,'' section in \textbf{SI}].

Having established that constructing a Lorentzian pulse is possible, we performed energy spectroscopy of the Fermi sea with the Lorentzian perturbation to demonstrate the minimization of electron-hole pair generation.

The idea was to apply a direct current (dc) on the upper-left ohmic contact, defining a voltage $V_L$, while the Lorentzian pulse is applied on the opposite right contact $C_I$. Under negative bias, electrons emitted in the energy range $-eV_L > \varepsilon > 0$, where $\varepsilon$ is energy, will antibunch with electron excitations coming from the driven right contact, resulting in no noise (see also Figs. S13 and S14). The noise variation with $V_L$ gives a measure of the number of electron excitations. The same procedure was repeated with positive bias to extract the number of hole excitations. Figure 2A shows the photoassisted shot noise as a function of the dc bias for single and two-harmonic modes (at $\varphi = 0$ and $\varphi = \pi$), which agrees well with the theoretical prediction for any injected charge $q$, which is an important requirement for the injection of Leviton ($q = 1$) and 2e-Leviton ($q = 2$). In Fig. 2, B and C, the excess shot noise $\Delta S_I$, obtained by subtracting the noise with $V_{\text{ac}}$ “off” from the noise with $V_{\text{ac}}$ “on,” is shown for a Lorentzian pulse constructed by summing four harmonics with $\alpha = q = 1$ (Leviton configuration) and $\alpha = q = 2$ (2e-Leviton configuration). The asymmetry of the excess noise reflects the absence of hole creation by the Lorentzian pulse [see also “Excess noise from sine and Lorentzian excitations" section in \textbf{SI}]. By comparing it to the computed excess noise for an ideal Lorentzian pulse (solid orange line), we experimentally verified that one or two electrons are injected along with a minimum amount of electron-hole pairs (at finite temperature thermal excitations add an extra contribution to the excess noise). This constitutes an experimental demonstration of on-demand single-electron injection in graphene quantum Hall edge channels. Note that for one- and two-electron Levitons, we did not observe any deviation from the non-interacting theory [see also “Relevance of the electronic interactions” section in  \textbf{SI}]. This measurement can be realized at different values of the transmission, which demonstrates the polar angle $\theta_1$ control of the 2e-Leviton state (Fig. 3).

\textbf{Coherent manipulation on the Bloch sphere}
We then turned on the lower beam splitter to showcase the coherent manipulation of single electrons as they propagate through the electronic MZI (Fig. 4, A and B). After passing the first beam splitter located at the upper edge of the PN junction, excited single electrons propagate along the N side with transmission probability $T_1 = |t_1|^2$ or the P side with reflection probability $|r_1|^2 = 1 - T_1$. In the Bloch sphere representation, the polar angle of the valley-isospin qubit is tuned by the upper side gate. When the polar angle is chosen to be $\pi/2$ (where the upper beam splitter is half-open as $T_1 = |t_1|^2 = 0.5$), the valley isospin of the initial state
\[
|\Psi_{\text{initial}}\rangle = \frac{1}{\sqrt{2}} \left(|\vec{w}\rangle + |\vec{-w}\rangle\right)
\]
at the entrance of the PN junction lies on the equator. The isospin then rotates around the $z$ axis by the azimuthal angle $\phi_{AB}$. The final state at the lower edge of the PN junction becomes
\[
|\Psi_{\text{final}}\rangle = \frac{1}{\sqrt{2}} \left(|\vec{w}\rangle + e^{i\phi_{AB}} |\vec{-w}\rangle\right).
\]
The value of $\phi_{AB}$ can be measured by the transmission probability $T_{\text{MZI}} = |\langle\tilde{\Psi}|\Psi_{\text{final}}\rangle|^2$, whereas the number of electron-hole pairs at the lower beam splitter can be measured by the shot noise.

We first applied this operation to sine-shaped voltage pulses. The shot noise is essentially determined by the total number of electron-hole pairs created by the pulse multiplied by the factor $T_{\text{MZI}}(1 - T_{\text{MZI}})$. It is given by two-particle scattering processes that enclose the AB-flux once (leading to a phase contribution $\phi_{AB} = 2\pi BA/\Phi_0$) or two times ($2\phi_{AB}$) [see also “Floquet scattering formalism for graphene Mach-Zehnder interferometry” section in \textbf{SI} for the full formula]. Figure 4, C to F, shows the shot noise as a function of the magnetic field for several frequency values ($\nu = 3.5, 7, 10.5, \text{and } 14 \, \text{GHz}$),
Investigated $\phi_{AB}$ values verifies that the key property of Leviton, namely the minimization of the number of electron-hole pairs with 2e injected charges, is conserved while propagating in the MZI. Furthermore, the value of $\phi_{AB}$ that is extracted from the amplitude of $\Delta S_I(q)$ is consistent with that obtained from $T_{\text{MZI}}$. These results demonstrate coherent control of Levitons.

\textbf{Conclusions and Outlook}
Our study demonstrates the emission and coherent control of a quantum state at the single-electron level in monolayer graphene. Although graphene interferometers have been studied in the dc regime \cite{jo2021,morikawa2015,wei2017, deprez2021,ronen2021} sending on-demand excitations or flying qubits toward a MZI is a notable step toward quantum information transfer. We established an on-demand electron source in graphene based on Levitons that minimizes the number of unwanted electron-hole pairs. By sending periodically excited Levitons to the MZI, we demonstrated the rotation of the valley flying qubit on the Bloch sphere. Encoding quantum information in the valley state of Levitons should enable two-valley qubit operations to be considered \cite{neder2007,samuelsson2004,ubbelohde2023,wang2023, fletcher2023}. There it can be shown that minimizing the number of electron-hole pairs is relevant against decoherence [see also “Levitons and decoherence”  section in (\textbf{SI}]. These on-demand electron pulses can also carry fractional charges, which offers the possibility to braid anyons \cite{nakamura2020,bartolomei2020, lee2020, roosli2021,lee2023} in graphene in the time domain. Graphene is emerging as a promising material with robust quantum properties compared with those of conventional semiconductors. Beam splitters, interferometers, and single-electron sources can be easily realized using PN junctions, opening up avenues for electronic quantum optics experiments. Owing to the simple and elegant circuit topology that exploits N and P counter-propagating edge states, complex, yet compact, interferometers with original entanglement schemes can be envisioned.

\section*{Acknowledgments}
We thank W. Dumnernpanich for his help with fabrication. 

\textbf{Funding:} This work was funded by European Research Council (ERC) starting grant COHEGRAPH 679531 (P.R.); by European Metrology Programme for Innovation and Research (EMPIR) project SEQUOIA 17FUN04, which is cofinanced by the participating states and the European Union’s Horizon 2020 program (P.R.); by the National Research Foundation of Korea via the SRC Center for Quantum Coherence in Condensed Matter (grant no. 2016R1A5A1008184 and RS-2023-00207732) (H.-S.S.); and by “Investissements d’Avenir” LabEx PALM (ANR-10-LABX-0039-PALM) (Project ZerHall) (F.D.P.). 

\textbf{Author contributions:} A.A., L.P., H.C., L.B., and P.R. performed the experiment with help from F.D.P.; A.A., L.P., H.C., L.B., N.K., S.L., H.-S.S., D.C.G., F.D.P., and P.R. analyzed and discussed the data; T.T. and K.W. provided the boron nitride layers; M.J. fabricated the device with input from A.A., F.D.P., and P.R.; P.R. wrote the manuscript with input from all co-authors; and P.R. designed and supervised the project. \textbf{Competing interests:} The authors declare no competing interests. \textbf{Data and materials availability:} Data and code are archived at Zenodo ( A. Assouline et al., Emission and coherent control of Levitons in
graphene. Zenodo (2023); https://zenodo.org/records/
10044265.)

\bibliographystyle{unsrt}
\bibliography{Ref}

\begin{figure*}
    \centering
    \includegraphics[width=\textwidth]{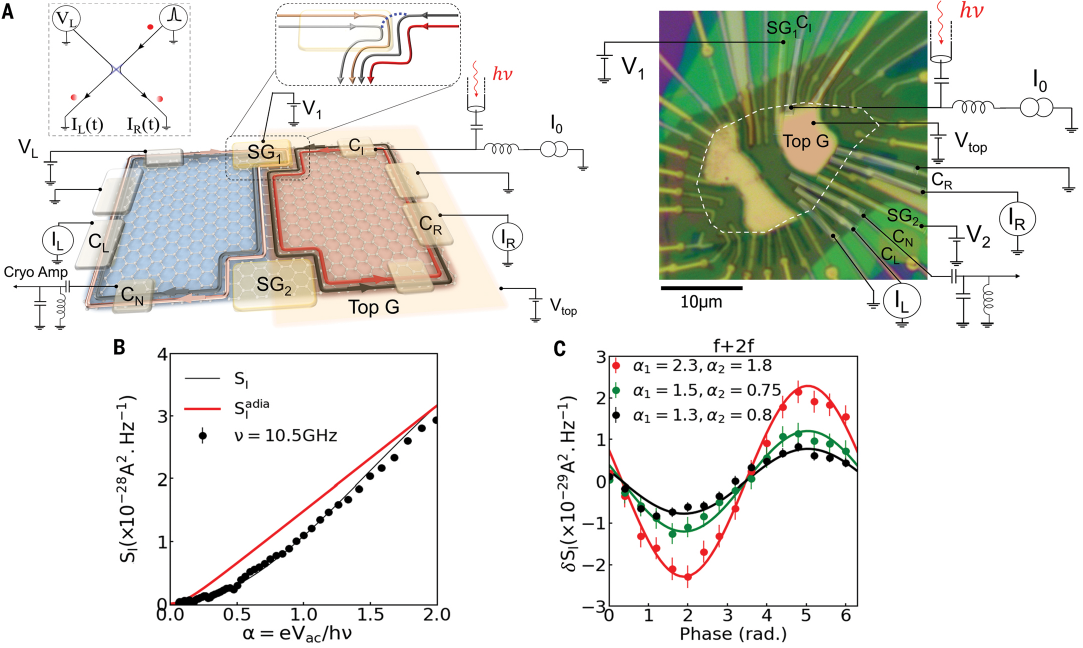} 
    \caption{
        \textbf{Photoassisted shot noise in the quantum Hall regime.} 
        \textbf{(A)} (Left) Schematic representation of the device in the bipolar quantum Hall regime. The filling factor below the top gate (Top G) is tuned to $\nu_P = -2$, whereas the region not covered by the top gate is tuned to $\nu_N = 2$. The filling factors below the upper and lower side gates (SG$_1$ and SG$_2$, respectively) are tuned to $\nu_1 \geq 2$ and $\nu_2 = 0$, respectively. 
        The voltage pulse at frequency $\nu$ is applied at the upper-right ohmic contact (C$_1$). Resulting photoassisted shot noise is amplified at low temperature (in contact C$_N$). Transmitted ($I_L$) and reflected current ($I_R$) are also measured. 
        The inset on the left shows a schematic representation of the partitioning experiment. The inset on the right shows a zoomed-in view of the valley splitter, where edge-state mixing occurs. (Right) Microscopic view of the sample obtained through an optical microscope. 
        \textbf{(B)} Measured shot noise $S_I$ (black dots) as a function of the photon number $\alpha$ at 10.5 GHz. The data are compared with predicted photoassisted shot noise (black line) [see also section IC, “Engineering voltage pulses,” in (32)] and the adiabatic (adia) excess noise (red line) given by 
        $S_I^\text{adia} = 2e \frac{2e^2}{h} D(1-D) \frac{1}{T} \int_0^T dt |V_\text{ac} \sin \left( \frac{2\pi t}{T} \right)| = 2e \frac{2e^2}{h} D(1-D) \frac{2V_\text{ac}}{\pi}$. 
        For $\alpha < 2$, the measured noise agrees with the predicted photoassisted shot noise. \textbf{(C)} Excess shot noise (circles) $\delta S_I = S(q=1, \alpha_1, \alpha_2, \varphi) - S(q=-1, \alpha_1, \alpha_2, \varphi)$ as a function of the phase difference between the first and second harmonics for different amplitudes ($\alpha_1 = eV_1 / h\nu$ and $\alpha_2 = eV_2 / 2h\nu$) compared with the predicted shot noise (solid lines) [see “Engineering voltage pulses,” section in \textbf{SI}]. 
        Error bars represent SEM.
    }
    \label{fig:photoassisted_noise}
\end{figure*}

\begin{figure*}[!ht]
    \centering
    \includegraphics[width=\textwidth]{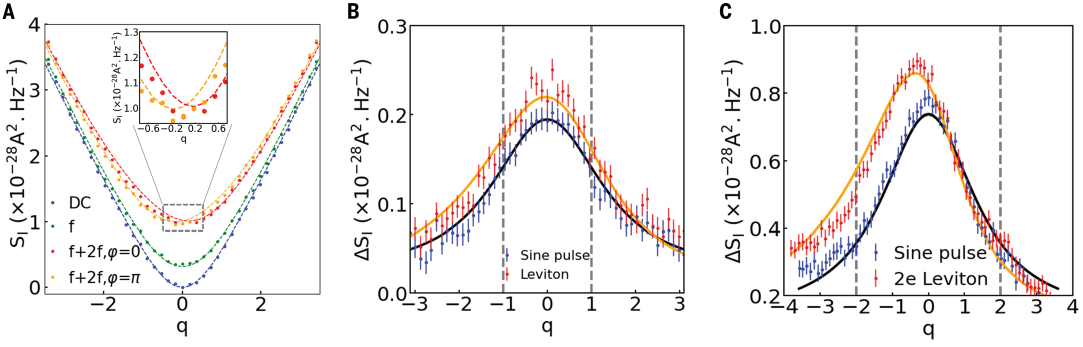}
    \caption{
    \textbf{Leviton and 2e Leviton.} 
    (A) Shot noise (circles) as a function of $q = eV_{\text{dc}} / h\nu$ for different values of $\alpha_1$, $\alpha_2$, and $\phi$ compared with the predicted shot noise (dashed lines) (\textbf{SI}). 
    Note that the finite temperature smears out the shot-noise singularities expected at $eV_{\text{dc}} = h\nu$. 
    The inset shows a zoomed-in view of the two-harmonics configuration. 
    (B) Comparison of the shot-noise spectroscopy for a 3.5-GHz sine wave with amplitude $\alpha = 1$ (blue circles) and a 3.5-GHz Lorentzian pulse with $W / T = 0.09$ ($W$ is the width of the pulse and $T = 1/f$, where $f$ is the pulse repetition frequency) and $\alpha = 1$ (red circles). 
    For the Lorentzian pulse, the excess shot noise $\Delta S_I$ is strongly asymmetric. For $q > \alpha + k_BT / h\nu$, the noise vanishes exponentially, which is the hallmark of the absence of hole excitations. 
    Solid lines are theoretical predictions (\textbf{SI}). 
    (C) Same as (B) with $\alpha = 2$. Note that the offset $\Delta q = -0.22$ may result from the existence of photocurrent at $\alpha = 2$. 
    We extracted an excess number of electron-hole pairs $\Delta N_{\text{eh}} = \Delta S_I / S_I^0 = 0.087$ for the Leviton and $\Delta N_{\text{eh}} = 0.076$ for the 2e Leviton. 
    In (B) and (C), the orange and black lines are the photoassisted shot noise theory for the Leviton and sine pulse, respectively, and error bars represent SEM.
    }
    \label{fig:leviton}
\end{figure*}

\begin{figure}[!ht]
    \centering
    \includegraphics[width=\columnwidth]{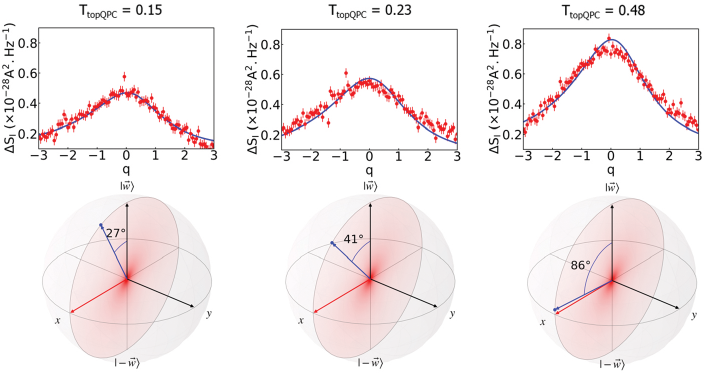} 
    \caption{
    \textbf{Polar angle control of the 2e Leviton state.} 
    Excess shot-noise measurements at three different values of the transmission for the top beam splitter of the MZI and corresponding polar angles in the Bloch sphere representation. 
    Error bars represent SEM. The solid blue line is the result of the photoassisted shot noise theory.
    }
    \label{fig:polar_angle_control}
\end{figure}

\begin{figure*}[!ht]
    \centering
    \includegraphics[width=\textwidth]{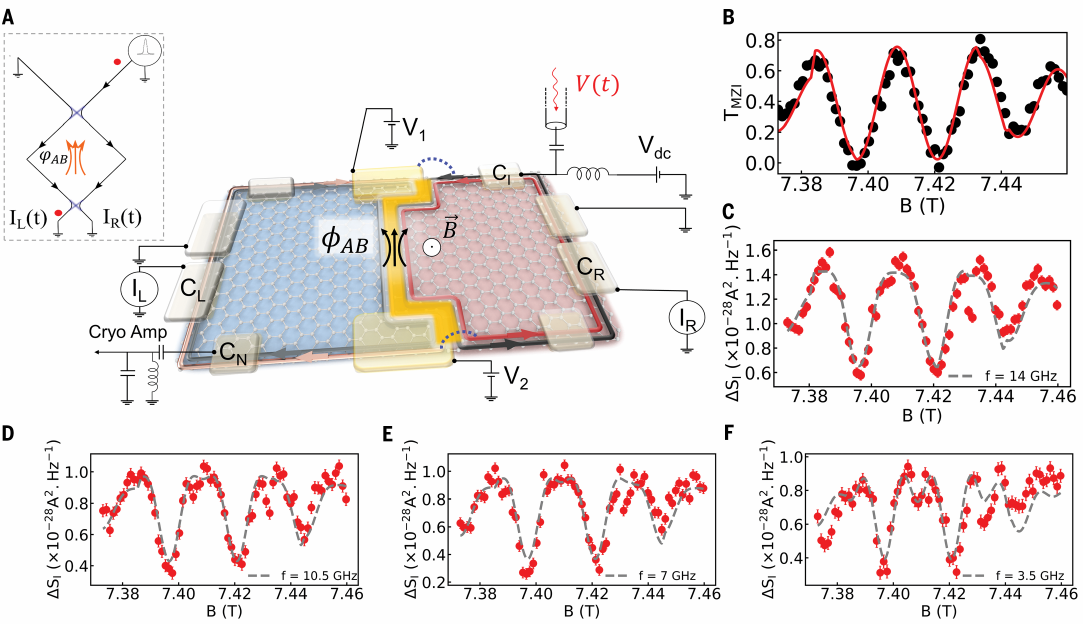} 
    \caption{
    \textbf{On-demand excitations in MZI.} 
    (A) MZI configuration with side gates tuned to $\nu_1 \geq 2$ and $\nu_2 \geq 2$. 
    The electron pulse accumulates a phase $\phi_{AB}$ upon propagation (the area enclosed by the loop is shown in yellow). 
    The inset shows a schematic representation of Levitons in an MZI experiment. 
    (B) Interference oscillations of the MZI transmission $T_{\text{MZI}}$ as a function of the magnetic field obtained by DC transport measurement. 
    The solid red line is the theoretical prediction, taking into account decoherence as a free parameter (10). 
    (C) Excess shot noise (red circles) generated by a sine wave propagating in an MZI as a function of the magnetic field for 14 GHz and $\alpha = 0.51$. 
    The dashed gray line is the theoretical prediction. [Detailed equations are given in “Floquet scattering formalism for graphene Mach-Zehnder interferometry” in section of \textbf{SI}.] 
    (D to F) As in (C) but for 10.5 GHz ($\alpha = 0.55$) (D), 7 GHz ($\alpha = 0.64$) (E), and 3.5 GHz ($\alpha = 1.22$) (F).
    }
    \label{fig:mzi_excitation}
\end{figure*}

\begin{figure*}[!ht]
    \centering
    \includegraphics[width=\textwidth]{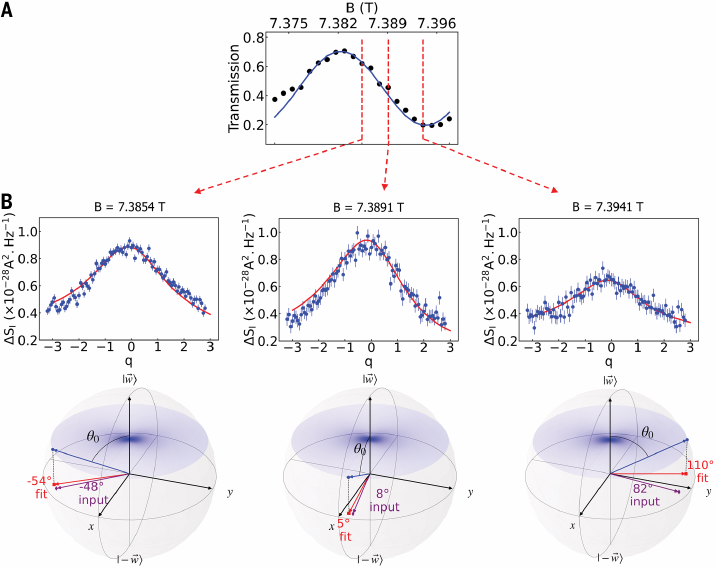} 
    \caption{
    \textbf{Coherent manipulation of a 2e Leviton.} 
    (A) MZI transmission $T_{\text{MZI}}$. The solid blue line is a sinusoidal fit used to extract the $\phi_{AB}$. 
    (B) (Top) Excess shot noise generated by a Leviton propagating in an MZI interferometer for different values of $\phi_{AB}$ ($B = 7.3854$, $7.3891$, and $7.3941$ T). 
    (Bottom) From left to right, the valley isospin rotates by an azimuthal angle of $\sim \pi$ on the Bloch sphere. 
    Fitted $\phi_{AB}$ (labeled “fit”) is compared to the actual value of $\phi_{AB}$ set by the magnetic field (labeled “input”). 
    The solid red line is fit to the photoassisted shot noise theory. 
    Error bars represent SEM.
    }
    \label{fig:coherent_2e_leviton}
\end{figure*}

\clearpage
\mbox{}


\section*{Supplementary material}

# I. MATERIALS AND METHODS:

## A. Experimental Setup

### 1. Sample fabrication

Graphene, hBN, and graphite flakes were mechanically exfoliated on a SiO2/Si substrate using the conventional scotch tape method. Graphene and graphite flakes were prepared by exfoliating bulk graphite from NGS Naturgraphit GmbH and hBN flakes were obtained by exfoliating bulk crystals from NIMS. A proper single-layer graphene flake was selected by optical microscopy, and its quality was further assessed after encapsulation with hBN by checking for any spurious signature in Raman spectroscopy [40]. All the crystals were carefully checked with optical microscopy and noncontact mode AFM for any defect, contamination and inhomogeneity in thickness before stacking. Using the van der Waals (vdW) dry transfer method, graphene was encapsulated between two hBN flakes[41]. The bottom hBN is 33nm thick, and the top hBN 27nm thick. This BN/graphene/BN stack was then transferred on a 9nm thick graphite flake, which is used as a back gate. The finished stack was annealed in vacuum at $\sim 350°C$ to clean up the sample. Bubbles and defects were checked and located by noncontact mode AFM. A polymethyl methacrylate (PMMA) pattern defined by e-beam lithography was used as a mask for O2/CHF3 reactive ion etch (RIE) to expose the 1D edge of graphene. Using the same PMMA layer as a liftoff mask, edge contacts were defined by e-beam evaporating Cr/Au=10/30nm while rotating the sample using a tilted rotation stage. We etched the graphene channel into the desired mesa geometry by using the same RIE procedure with PMMA resist mask. PMMA resist was cleanly removed in hot acetone. To prevent any possible leakage between the exposed graphene edge and additional gates, a 10nm thick hBN flake was picked up and dry transferred on top of the sample. This time, annealing wasn't used

3

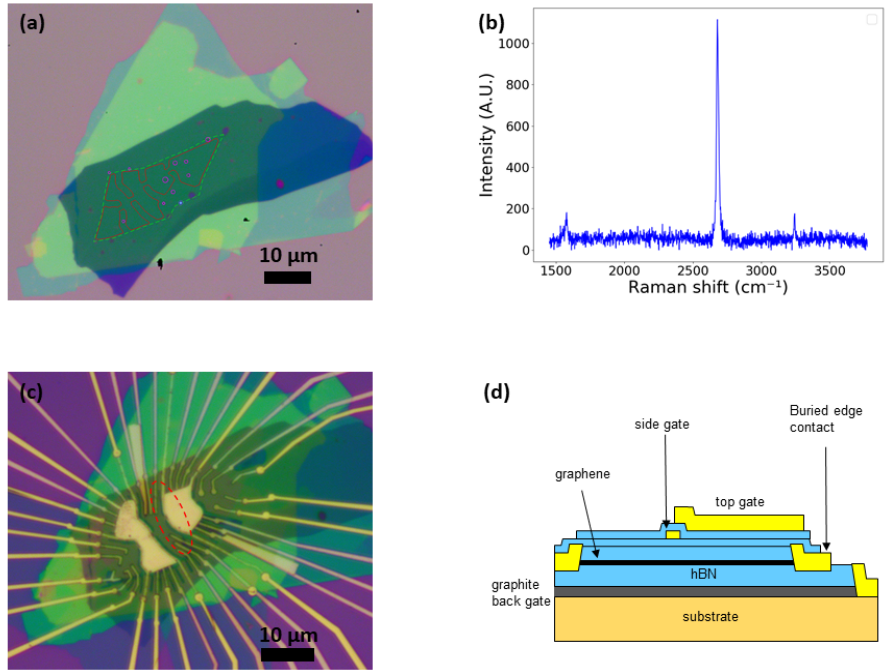

Figure S 1: **Sample description**

**A**: Optical image of (top to bottom) hBN/graphene/hBN/graphite stack. The green dashed line indicates the edge of graphene. Red line indicates graphene mesa which would be defined by subsequent RIE etch. Purple circles indicate bubbles in the stack which later would be eliminated by ohmic contacts fabrication or mesa etching. **B**: Raman spectrum of encapsulated single-layer graphene. This data was taken after successful encapsulation of graphene with top and bottom hBN but before adding additional graphite. We could minimize the spurious substrate effect and also avoid overlapping with the spectrum of graphite underneath. **C**: Optical image of a finalized device. The dashed red line indicates one PN-junction, which is defined by applying gate voltages from the graphite back gate and metallic top gate. **D**: Cut plane view of the sample, equivalent of dashed red line section in **C**

to remove the resist (PPC) used for the transfer, to avoid damage to the ohmic contacts. It was instead cleaned with hot acetone. Side gates were defined by evaporating Cr/Au=10/30nm. An additional 15nm thick hBN flake was picked up and dry transferred on top of the sample like described before. Finally, the top gates were defined by evaporating Cr/Au=10/40nm (in Fig. S1).

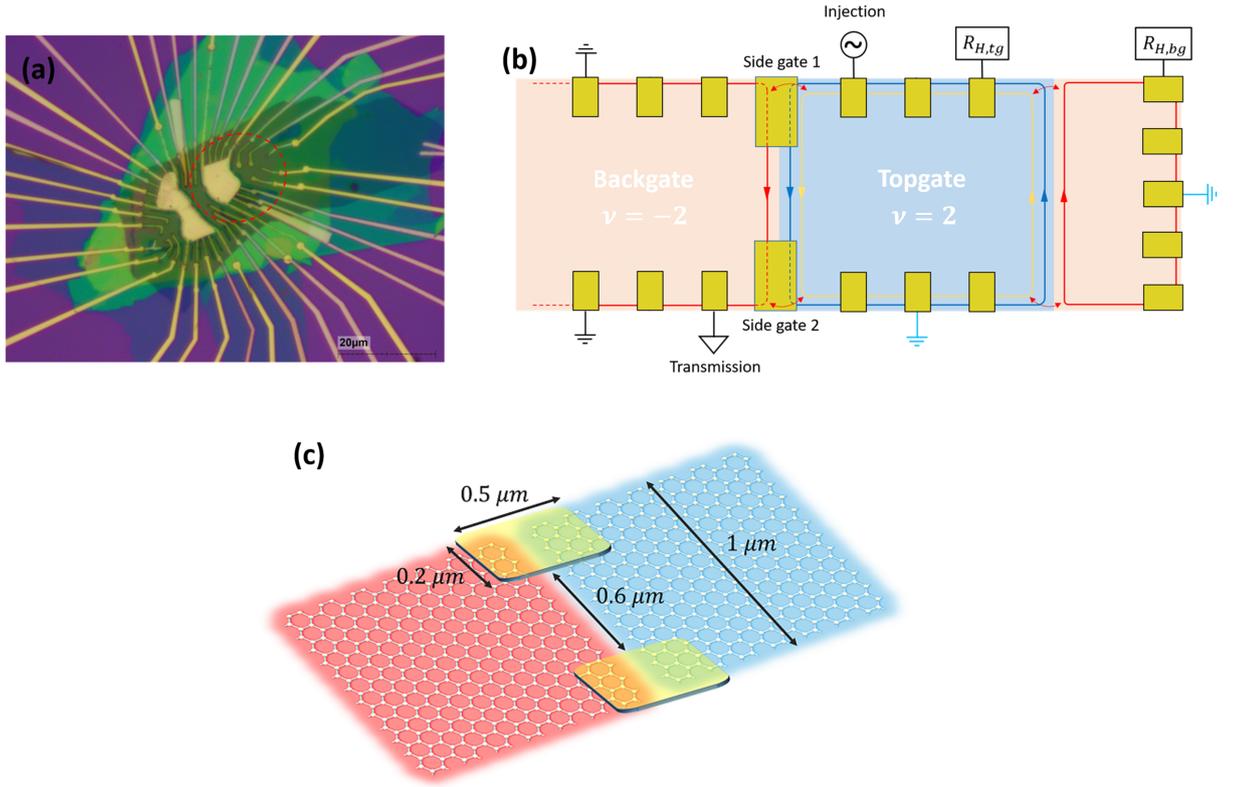

Figure S 2: **Sample geometry.**

**A**: Optical micrograph of the finalized sample composed of two top gates (corresponding to the two golden plates in the center). The top gate used in this study is circled by a red dashed line. Two PN junctions are generated by the top gate : one on the right without side gates plays the role of a test sample, and another one on the left with side gates corresponds to the valley splitter sample. **B**: Schematic of the used part of the sample (circled by the red dashed line in **A**). The n-doped region (blue) is set at filling factor $\nu = 2$. The p-doped region under the top gate(red) is at $\nu = -2$. On the left junction, current was injected from the top right contact, and the transmitted current was measured through the bottom left contact. Light blue grounds correspond to cold grounds. The two-point Hall resistance was measured under the top gate (denoted $R_{H,tg}$) and outside (denoted $R_{H,bg}$). **C**: Characteristic dimensions of the valley splitter sample : the sample width $1.0 \mu m$, the distance between the side gates $0.6 \mu m$, and a total side gate length $L_s$=450nm

2. *Sample geometry and RF set up*

The fabricated sample includes two top gates (in Fig. S2a). Only the right top gate circled by a red dashed line has been used in this study. As shown in the schematic of the used part of the sample (in Fig. S2b), two PN junctions are generated by the top gate : the one on the right without side gates plays the role of a test sample, and the other one on the left with side gates

5

corresponds to the valley splitter sample. Precise dimensions of the valley splitter sample are presented in Fig. S2c. The width of the graphene ribbon is $1.0 \mu m$. The side gates are rectangles of length $0.20 \mu m$ and width $0.5 \mu m$ separated by distance $0.6 \mu m$. Half of the side gates is below the top gate, which implies that the PN junction meets the side gates borders in their middle at a distance $0.25 \mu m$ of the two other borders of each side gate. Electrons propagate along the edge of the side gate over a length $L_s$=450 nm. The total PN interface length is $2L_s+0.6=1.5$ $\mu$m as mentioned in the main text.

The sample is connected to the coaxial lines via a Printed Circuit Board (PCB). Each coplanar line connects the coaxial lines on one end and the ohmic contacts of the sample on the other end. The rf-line connecting the room temperature Sine Wave Generator (Anapico synthesizer 300KHz-20GHz) to the sample is designed in order to give very low loss. Between 300K-4K where the dissipation may be significant we use ultra-low loss SC-219/50-SB-B DC-40GHz (KMCO) cryogenic coaxial lines. From 4 K to the lowest temperature we have superconducting coaxes NbTi DC-40GHz (KMCO) to minimize heating of electrons at base temperature. To prevent thermal microwave photons from the hot part of the rf-circuit to reach the sample, attenuators are regularly distributed at the different temperature stages of the refrigerator for a total of ∼50dB.

3. *Shot Noise set up*

<u>Resonant RLC tank circuit:</u>

The voltage fluctuations of a RLC tank circuit are measured to extract the current noise of the sample. The RLC tank circuit is composed of the sample resistance $R_s$, in parallel with an inductor $L$ (22 $\mu H$), a capacitor $C$ (240 $pF$) and a resistor $R_p$ (20 $k\Omega$). Its resonance frequency is centered at 2.2 $MHz$. The capacitance $C$ is mostly due to the distributed capacitance along the coaxial line connecting the sample to the cryogenic amplifier. About one meter of this line is wrapped with silver epoxy around the cold body of the fridge at 10mK, together with all the DC connections to the sample, in order to ensure a low electronic temperature. The dissipation in the circuit is due to the DC resistance of the inductor (about 10 $\Omega$ at room temperature), which is taken into account by adding a resistance in series with the inductor. The resistance $R_p$ is used to adjust the resonance bandwidth and to shunt the spurious signal coming from the 4K amplifier. Additional blocking capacitors are used to prevent the DC current from the sample to enter the RLC tank circuit.

Amplification chain:

The voltage fluctuations of the RLC tank circuit are first amplified using a low noise amplifier, based on a commercial HEMT (ATF-34143), placed on the 4K plate of the dilution fridge.

After a second stage of amplification at room temperature (amplifier NF SA-421F5), the signal is digitized with an acquisition card (adlink) to compute the noise spectral density.

In Fig. S3A, the voltage noise spectrum is shown for different temperatures. Note that the voltage spectral density at 20 mK is subtracted from all curves to remove the noise background. Each curve is fitted to extract the gain of the amplification chain (G ≈ 800).

Base sample temperature:

As shown in Fig. S3B, the current noise of the circuit is quasi-linear in temperature down to 25 mK as described by the Johnson-Nyquist formula. Therefore, we consider that the sample electronic temperature equals the fridge temperature when no voltage drive is applied to it.

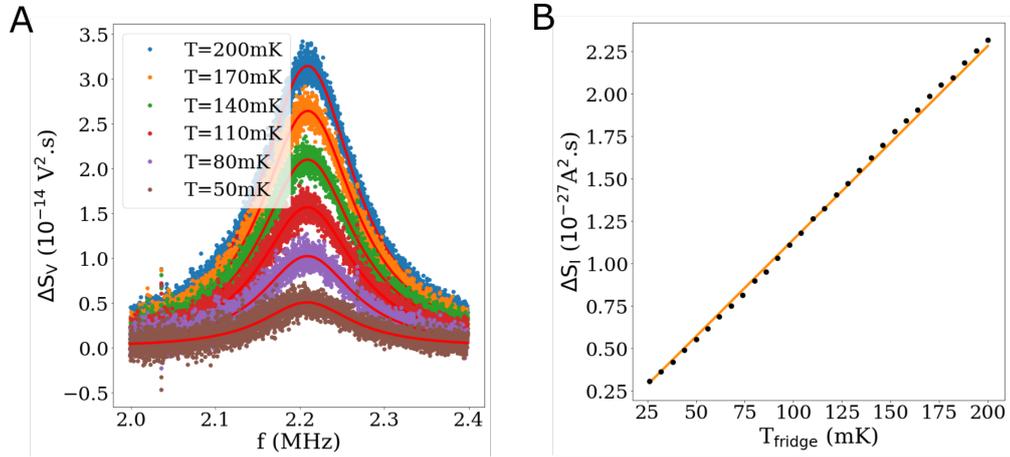

Figure S 3: **Calibration of the gain of the amplification chain**
**A**: Voltage noise spectrum for different temperature after amplification. The red curves are the fits of the Johnson-Nyquist noise of the circuit which is composed of the sample resistance in parallel with a RLC resonator. The gain of the amplification chain is obtained from the fit. **B**: Current noise of the circuit as a function of temperature.

**B. DC shot noise and photo-assisted shot noise**

*1. Transmission dependence of the noise*

Using the Landauer-Büttiker formalism, the excess current noise induced by the partitioning of several modes through a scatterer is given by:

7

$$\Delta S_I = \frac{2e^2}{h} \sum_n D_n(1 - D_n) \left[eV_{dc} \coth\left(\frac{eV_{dc}}{2k_B T}\right) - 2k_B T\right] \quad (1)$$

where n indexes electronic channels, $D_n$ is their transmission, $V_{dc}$ is the voltage bias, and T is the electronic temperature.

In the quantum Hall regime, the number of channels is tuned by the filling factor, and the transmission on both edges of a pn junction can be adjusted using side gates, as described previously [42]. In Fig. S4A the transmission D of the upper edge of the pn interface is tuned using the top side gate while the mixing on the bottom edge is cancelled by fixing the bottom side gate value to V=-0.5V. The shot noise shown in Fig. S4B for different transmissions is in agreement with the scattering theory for two channels. This confirms unambiguously that both channels are mixing at the edge of the pn junction [43]. The Fano factor of the top and bottom beam splitter is extracted in Fig S4C and D respectively.

2. *Wiedemann-Franz heating*

Injection of current through an ohmic contact is expected to cause some heat dissipation in the sample. Starting from the heat equation, we use an analogue of the Wiedemann-Franz equation $\frac{\kappa}{\sigma} = \beta \cdot T$ to frame the problem in terms of the electrical conductivity. After integration of the heat equation [44], the heating model becomes : $T(V) = \sqrt{T_0^2 + \beta \langle V^2 \rangle}$, where $T_0$ is the base temperature of the sample when no electrical excitation is provided, and $\beta = \frac{6e^2}{\pi^2 k_B^2} \frac{RDG_0}{4} \left(1 + \frac{RDG_0}{2}\right)$. R is the resistance of the ohmic contact, D the valley splitter's transmission and $G_0 = \frac{2e^2}{h}$. Due to the different paths that the DC and RF excitations could take through the sample, and in order to account for possibly different behavior of the ohmic contacts in the RF range, we consider a generalized version of this heating model : $T(V_{DC}, V_{AC}) = \sqrt{T_0^2 + \beta_{DC} V_{DC}^2 + \beta_{AC} \frac{V_{AC}^2}{2}}$ that treats $\beta_{DC}$ and $\beta_{AC}$ as fitting parameters. Figures S5, S6 and S7 show the effects of heating on the current noise with respect to DC bias and RF excitations. Fitting gives a $\beta_{DC}$ value between $10^6 K^2 \cdot V^{-2}$ and $1.2 \cdot 10^7 K^2 \cdot V^{-2}$. From the fitting of both $\beta_{DC}$ and $\beta_{AC}$ from a noise measurement, it is possible to assess the impact of AC heating from RF bias calibrations.

3. *Photo-assisted shot noise at $\nu=7$ GHz*

Additional data of the photo-assisted shot noise induced by a 7 GHz sine excitation are shown in Fig. S8.

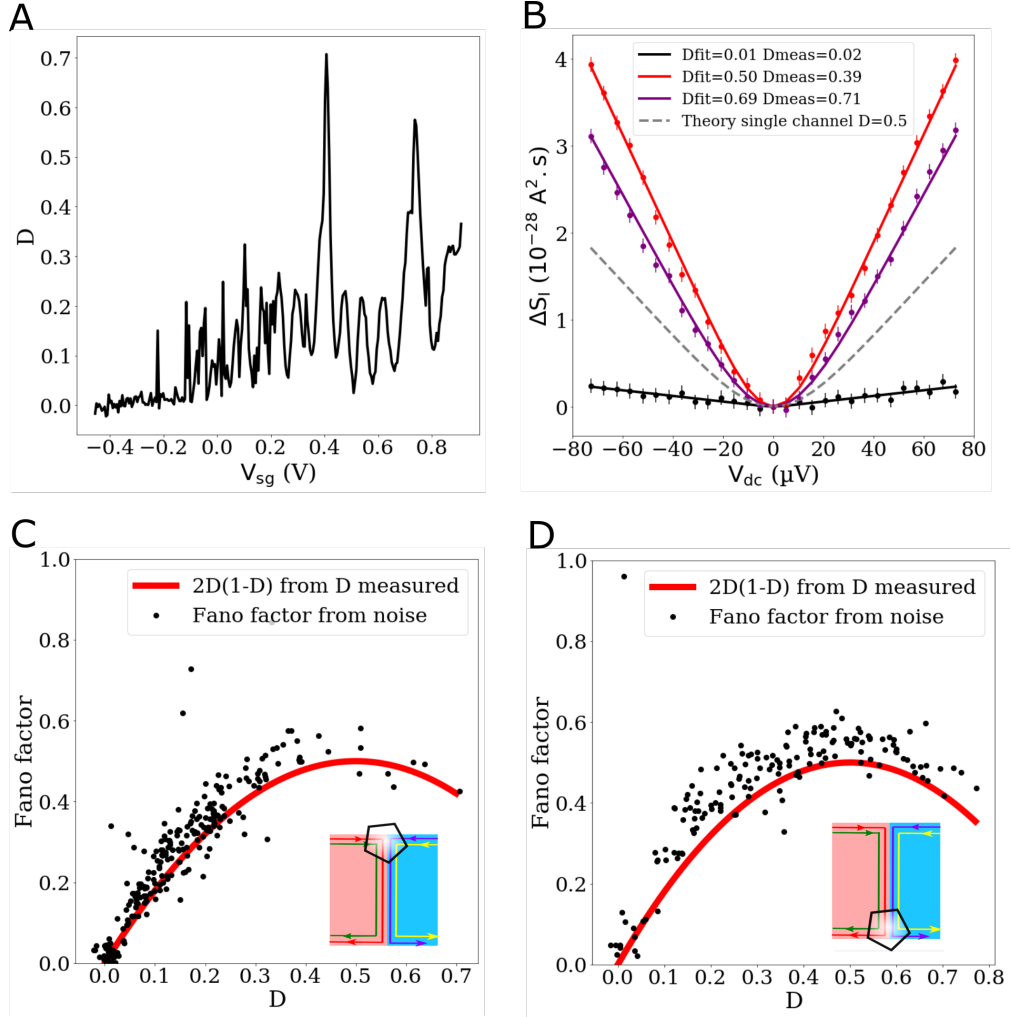

Figure S 4: **Transmission dependence of the noise**

**A**: Transmission D as a function of top side gate voltage. The bottom side gate voltage is fixed at -0.5V to avoid the mixing at the bottom beam splitter. **B**: Current noise as a function of DC voltage for different transmissions. The measured transmission is in good agreement with the transmission obtained by fitting the noise with the scattering formula for two channels. Grey dashed line shows the maximum shot noise obtained from a single channel which can not account for the experiment. **C**: Noise fano factor of the top beam splitter for the different transmissions D shown in **A** (92 percent of the data points can be fitted). In red the calculated 2D(1-D) obtained from the transmission measurement. **D**: Noise fano factor as a function of transmission D for the bottom beam splitter. Here the top side gate is fixed to -0.5V to avoid mixing at the top beam splitter.

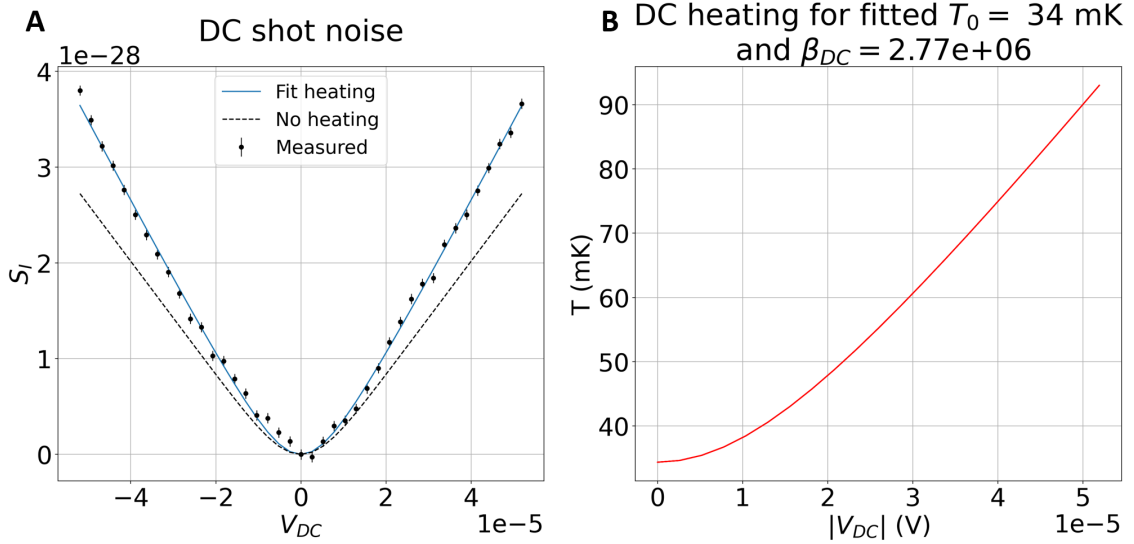

Figure S 5: **Wiedemann-Franz heating for a DC excitation in a graphene PN junction**
**A**: Effect of the electron heating on the DC shot noise. Adding the heating term to the temperature causes an increase in the absolute slope $\frac{\partial S_I}{\partial V}$ of the DC shot noise curve, which could also be interpreted as a higher Fano factor. **B**: Evolution of the electronic temperature as a function of the DC bias with the $\beta_{DC}$ coefficient fitted from the measurement in **A**.

### C. Engineering voltage pulses

#### 1. "In situ" control of the phase and the amplitude of a pulse

In the following, we detail how we can control "in situ" the phase and the amplitude of an injected pulse in the sample. Following Floquet scattering theory, when an electron from the right reservoir experiences an arbitrary periodic potential $V_{ac}(t)$ of frequency $\nu$, its wave function gains an extra phase term $\varphi(t) = 2\pi \frac{e}{h} \int_{-\infty}^{t} Vac(t')dt'$. Developing this new $e^{-i\varphi(t)}$ term into its Fourier series $\sum_l p_l \cos(2\pi l \nu t)$, with Fourier amplitudes $p_l$, gives a resulting state that is a superposition of electronic states with energies shifted by $lh\nu$ and probability amplitudes $p_l$. This can be seen as probabilities for the electron to absorb ($l > 0$) or emit ($l < 0$) photons, hence photo-assisted shot noise. The resulting electron-hole pairs are then partitioned at the valley splitter generating photo-assisted shot noise $S_I^{PASN}$. At zero voltage bias $V_{dc} = 0$ and zero temperature $T_e = 0$, this gives $S_I^{PASN} = S_I^0 \sum_{l=-\infty}^{\infty} |l| P_l$ with the probability $P_l = |p_l|^2$. The measurement of $S_I^{PASN}$ is therefore directly proportional to the total number of excited electron-hole pairs. In order to obtain a single electron source, a finite voltage bias $V_{dc}$ has to be applied with $\alpha = \frac{eV_{ac}}{h\nu} = q = 1$. The relevant quantity to extract the number of electron-hole pairs that are emitted in addition to the single elec-

10

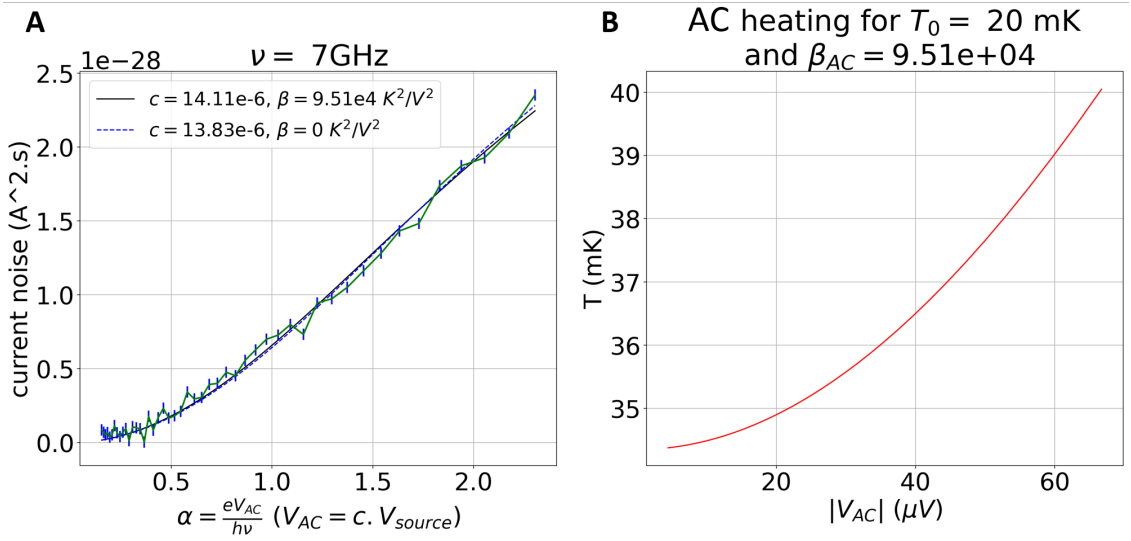

Figure S 6: **Wiedemann-Franz heating for a 7 GHz AC excitation in a graphene PN junction**
**A**: Calibration curve (input amplitude vs. shot noise) for a 7 GHz excitation. Here, the difference between the noise with and without a fitted $\beta_{AC}$ to take heating into account is negligible. **B**: Evolution of the electronic temperature as a function of the AC bias with the $\beta_{AC}$ coefficient fitted from the measurement in **A**.

tron is the excess noise $\Delta S_I = 2\frac{2e^2}{h}D(1-D)\sum_{l=-\infty}^{\infty} P_l|lh\nu + eV_{dc}|$. Experimentally, the electronic temperature is finite and the expression can be rewritten $\Delta S_I(V_{ac}, V_{dc}, T_e) = S_I^0 \Delta N_{eh}(\alpha, q, \theta_e)$ with $\Delta N_{eh}(\alpha, q, \theta_e) = \sum_{l=-\infty}^{\infty} (l+q)coth(\frac{l+q}{2\theta_e})P_l(\alpha) - q \times coth(\frac{q}{2\theta_e})$. In Fig. S9**B**, $S_I^0 N_{eh}(\alpha, q, \theta_e)$ (with $N_{eh}$ the total number of electron-hole pairs) is plotted as a function of q for four different configurations: the DC transport shot noise $S_I^0 N_{eh}(\alpha = 0, q, \theta_e)$ the photo-assisted shot noise with a sine excitation at 3.5GHz $S_I^0 N_{eh}(\alpha = 1.09, q, \theta_e)$, and the biharmonic case with two phases $\varphi=0$ and $\varphi=\pi$ corresponding to $S_I^0 N_{eh}(\alpha_1 = 1.09, \alpha_2 = 1.1, q, \varphi = 0)$ and $S_I^0 N_{eh}(\alpha_1 = 1.09, \alpha_2 = 1.1, q, \varphi = \pi)$. The asymmetry between the two phase configurations $S_I^0 N_{eh}(\alpha_1 = 1.09, \alpha_2 = 1.1, q, \varphi = 0)$-$S_I^0 N_{eh}(\alpha_1 = 1.09, \alpha_2 = 1.1, q, \varphi = \pi)$ is confirmed (plotted in Fig. S9C). We proceed similarly for the phase calibration between the first and the third harmonics, by measuring the shot noise as a function of the phase difference between the first and the third harmonic (in Fig. S10).

2. *Excess noise from sine and Lorentzian excitations*

Providing RF excitations to the sample leads to photo-assisted phenomena, creating excess noise that adds to the DC shot noise. If a signal is written as $V_{ds}(t) = V_{DC} + f(t)$, and f(t) is sinusoidal, the excess shot noise is symmetrical under flipping the sign of $q = \frac{eV_{DC}}{h\nu}$, whereas it is

11

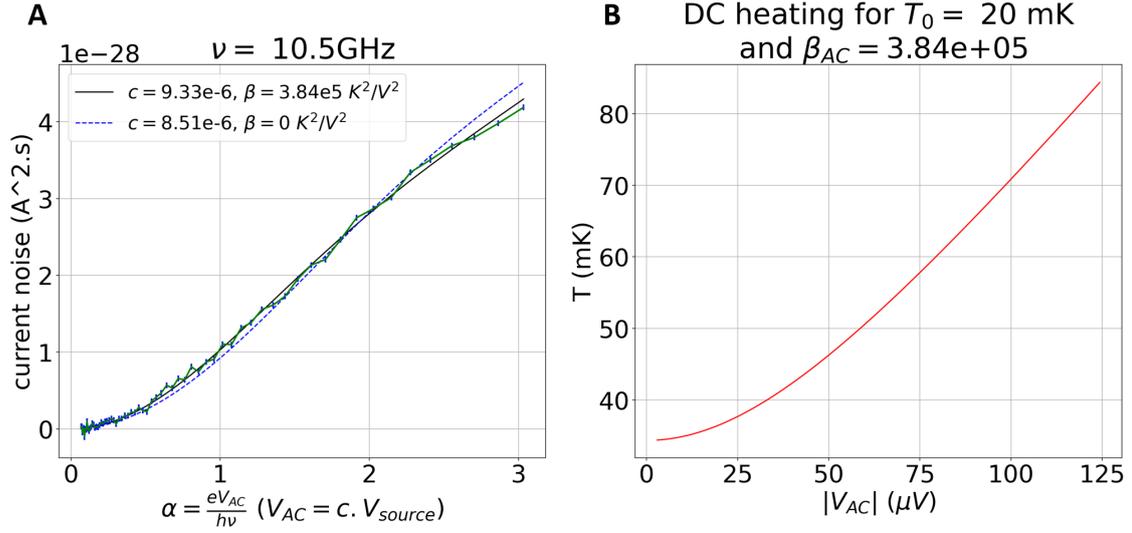

Figure S 7: **Wiedemann-Franz heating for a 10.5 GHz AC excitation in a graphene PN junction**
**A**: Calibration curve (input amplitude vs. shot noise) for a 10.5 GHz excitation. At higher frequencies, fitting the heating visibly corrects discrepancies between the measurement and the fit without heating. Still, 10.5 GHz is the third harmonic of our Leviton, for which we apply a $\alpha$ well below 1, in a region where the AC heating is, again, negligible. **B**: Evolution of the electronic temperature as a function of the AC bias with the $\beta_{AC}$ coefficient fitted from the measurement in **A**.

not for Lorentzian pulses. Figures S11 and S12 show that at zero temperature, the excess shot noise generated is piecewise linear, whereas nonzero temperatures give a shot noise curve that is gradually smoothed out. Notably, in the case of a Lorentzian excitation and $\alpha$=1, the excess noise vanishes at $q = 1$ the hallmark of the absence of electron-hole pairs. This defines a pure electronic excitation without unwanted electron-hole pairs. Figure S13 shows a noise measurement displaying these characteristics, in the case of a 2e Leviton, at $\alpha = 2$, as well as the corresponding distribution function. The left figure depicts the effect of a Lorentzian pulse on the Fermi distribution. Electrons from the Fermi sea might have absorbed (l > 0) or emitted (l < 0) l photons with probability $|P_l|$, generating partial upward or downward shift of the Fermi sea. Energy population for the out-of-equilibrium conductor is given by the modified distribution $\tilde{f}_L = \sum_{l-\infty}^{+\infty} P_l f_l(\epsilon - lh\nu)$. More precisely, $\tilde{f}_L$ is asymmetrical because no hole is created, which is a hallmark of Levitons. For a 2e-Levitons, this is equivalent to $|P_l|$=0 for l<-2. This property stems from the fact that a dc voltage $V_L$ is also applied with $V_L$=Vac=2h$\nu$/e. Without any dc voltage, we would find $|P_l|$=0 for l<0. The excess shot noise measurements, that act as an energy spectroscopy, reflects the modified Fermi distribution and will also be asymmetric as a function of the dc voltage q=eV$_R$/h$\nu$ applied on the opposite contact. The energy spectroscopy of the modified Fermi sea works as follow (see also

12

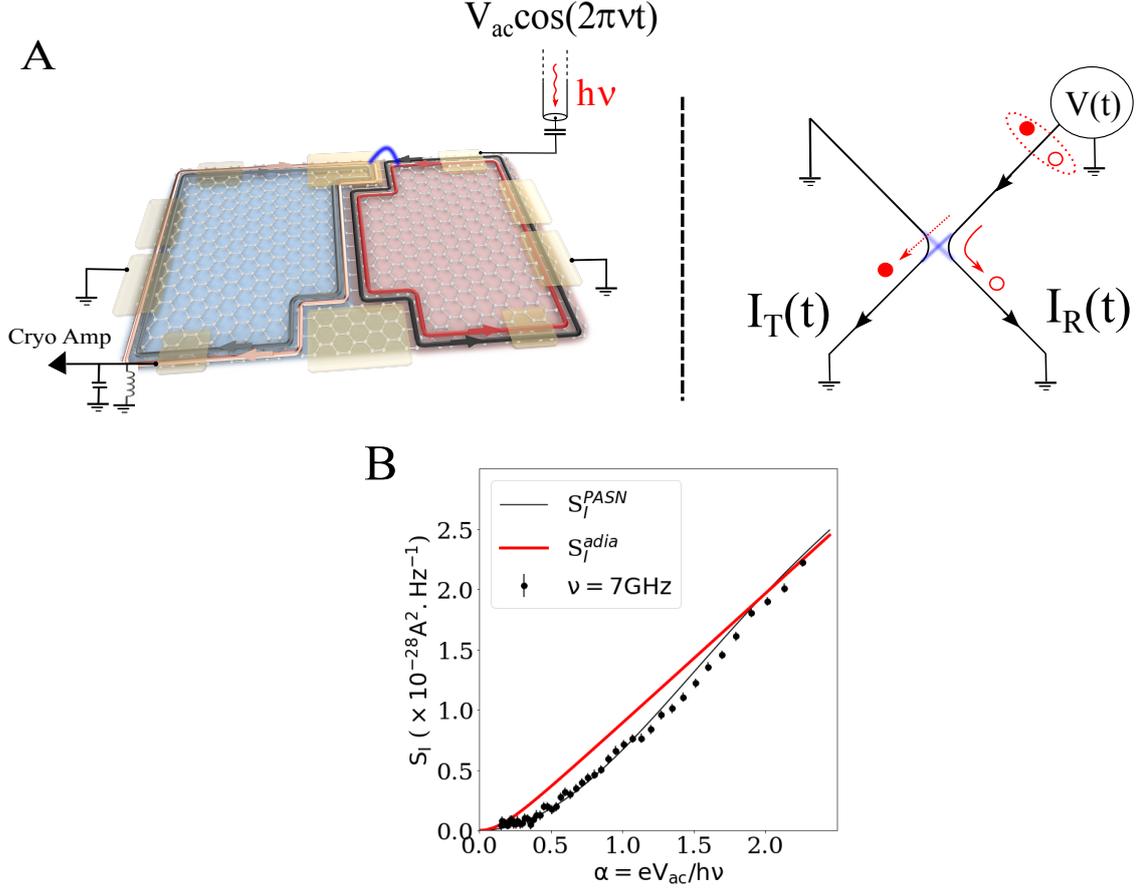

Figure S 8: **Photo-assisted shot noise in the quantum Hall regime**

**A**: Schematic representation of the device in the quantum Hall regime. A metallic top gate enables tuning the filling factor below this latter to $\nu_T$=-2. Filling factor of the region which is not covered by the top gate is tuned to $\nu_B$=2. Filling factor below the top side gate is tuned to $\nu_1 \geq 2$, and to $\nu_2 = 0$ below the bottom side gate. The voltage pulse at frequency $\nu$ is applied at the top right ohmic contact. Resulting photo-assisted shot noise is amplified at low temperature. Right: Schematic representation of the experiment. The ac excitation V(t) generates electron-hole pairs which are then partitioned at the valley splitter. Transmitted and reflected current as well as current fluctuations are measured. **B**: Measured photo-assisted shot noise $S_I^{PASN}$ (black dots) as a function of the photon number $\alpha$ at 7 GHz. The experimental noise agrees with the predicted photo-assisted shot noise (black solid line). In red solid line, theoretical excess noise in the adiabatic limit.

S14): a direct current (DC) is applied on the right ohmic contact, defining a voltage $V_R$, while the Lorentzian pulse is applied on the opposite left contact. Under negative bias $V_R < 0$, electrons emitted in the energy range $-eV_R > \epsilon > 0$ will anti-bunch with electron excitations coming from the driven left contact, resulting in no noise. By measuring the noise variation as a function of $V_R$, we can extract a precise number of electron excitations. Same procedure is repeated with positive

13

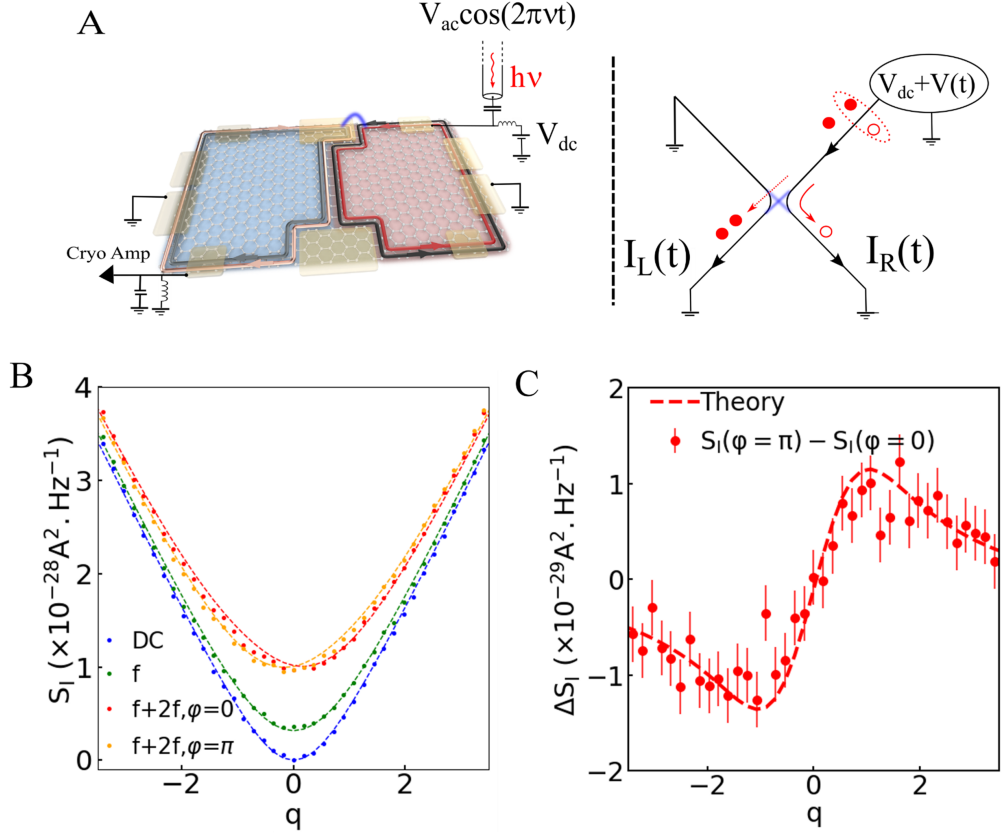

Figure S 9: **Engineering of an electron source**

**A**: Schematic representation of the electron source. A DC source in parallel to an AC excitation are used. Right: Injected electrons is accompanied by unwanted electron-hole pairs. The number of electron-hole pairs depends on the voltage pulse shape V(t). **B**: Shot noise as a function of q= $eV_DC/h\nu$ for different excitations. Blue: no RF excitations. Green: sine pulse. Red: combination of the first and second harmonics. Orange: combination of the first and second harmonics with a phase difference between these two equal to $\varphi=\pi$. (C) $S_I(\alpha_1 = 1.09, \alpha_2 = 1.1, q, \theta_e, \varphi = \pi)$- $S_I(\alpha_1 = 1.09, \alpha_2 = 1.1, q, \theta_e, \varphi = 0)$ as a function of q.

bias $V_R > 0$ to extract the number of hole excitations. Therefore, in the absence of hole excitations the shot noise will be zero.

3. *Description of a truncated Leviton*

Getting an exact Lorentzian profile $V_p(t) \propto \sum_{k=-\infty}^{\infty} \frac{1}{1+\left(\frac{t-kT}{W}\right)^2}$ [45] from an arbitrary waveform generator (AWG) is not easy to achieve because the nonuniform frequency response of the setup

14

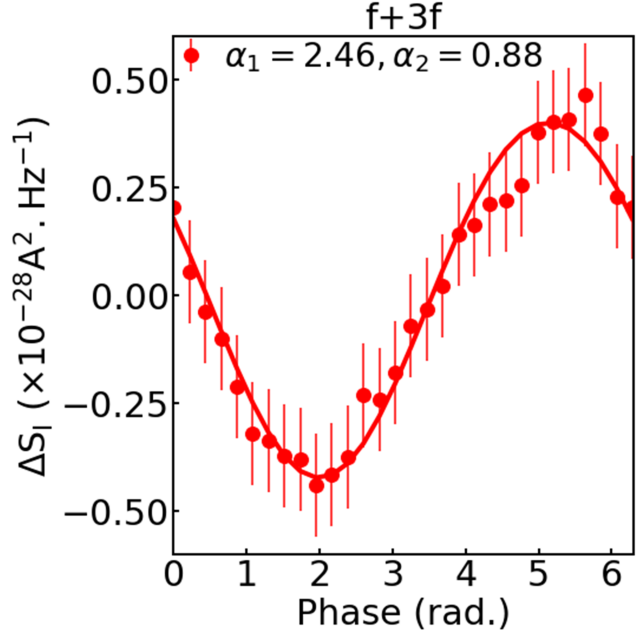

Figure S 10: **Relative phase calibration between $f = 3.5GHz$ and $3f = 10.5GHz$ harmonics** Excess shot noise plotted as a function of the phase difference between the first harmonic and the third one for different amplitudes ($\alpha_1 = eV_1/h\nu$ and $\alpha_3 = eV_3/3h\nu$). Such measurement, when compared to photo-assisted theory, allows to set the relative phase between any two harmonics.

deforms the pulse as it propagates through the RF lines. Instead, the time domain representation of the Lorentzian pulses is decomposed into a Fourier series : $V(t) \propto 1 + 2\sum_{l=1}^{+\infty}\cos(2\pi\nu l t)e^{-2\pi l \frac{W}{T}}$ [46]. Each harmonic of this pulse can then be controlled independently and adjusted to compensate their respective dephasing through the lines. The pulse is then approximated by truncating the infinite series to a finite one. The contribution of each harmonic to the Lorentzian profile is exponentially decreasing and is dictated by the width parameter $\frac{W}{T}$, chosen here to equal 0.09 to give a sufficient trade-off between lower noise (which is achieved with lower widths, i.e. sharper peaks) and acceptable convergence when doing few-harmonic reconstitution (which requires lower $\frac{W}{T}$). The effect of this truncation of the sum is shown in fig. S15. At $\frac{W}{T} = 0.09$, a three-harmonic reconstruction is sufficient to faithfully reproduce the distribution of the probabilities $P_l$ of shifting the Fermi sea by $lh\nu$, including finer features visible on the right part of the plots and up to high electron-per-pulse injection numbers. Fig. S16 shows how adding more harmonics influences both the value and the asymmetry of the excess photon-assisted shot noise in the case $\frac{W}{T} = 0.09$, showing fast convergence towards the value expected of a complete Lorentzian profile.

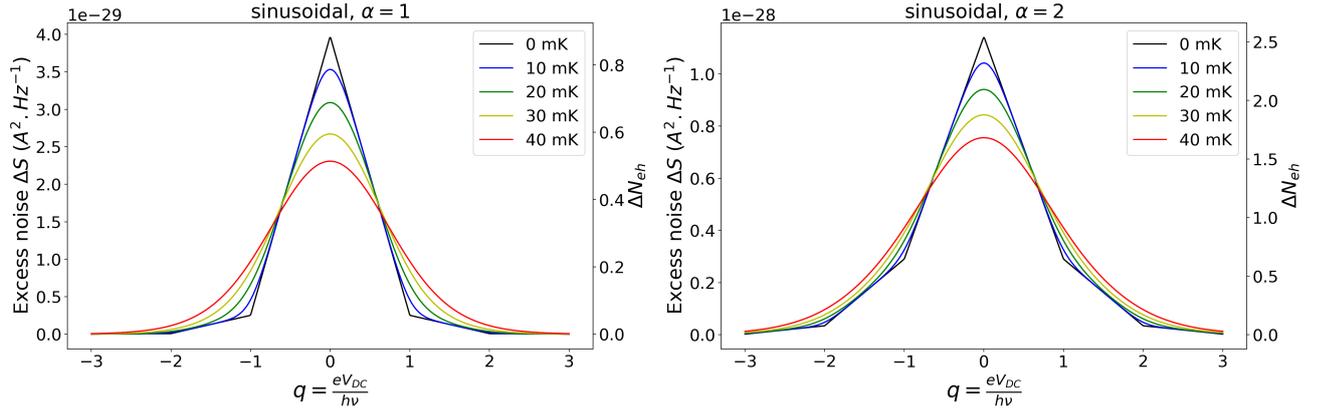

Figure S 11: **Excess photo-assisted shot noise of sine waves**

Excess photo-assisted shot noise with sine excitations of reduced amplitude $\alpha = \frac{eV_{AC}}{h\nu}$ for different values. Nonzero temperatures induce smearing of the curve.

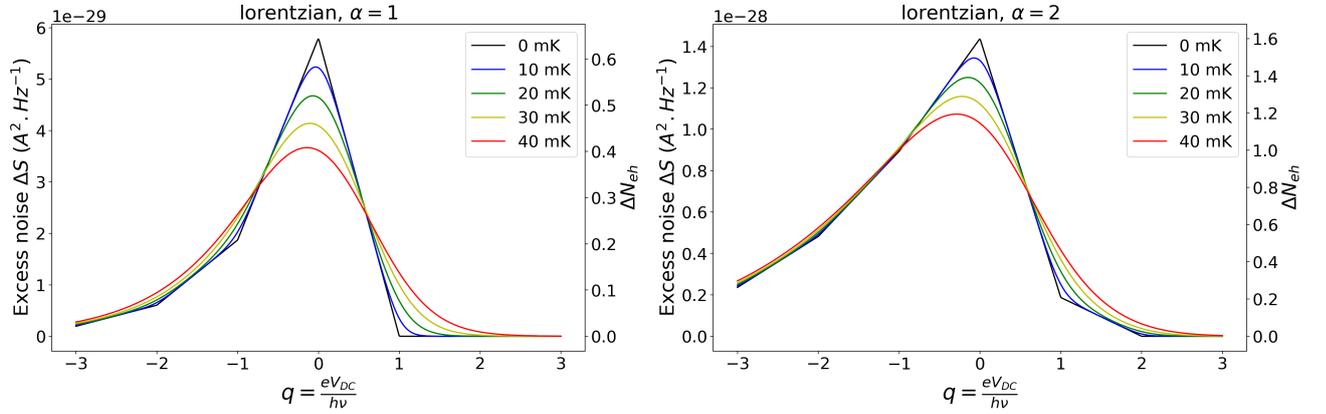

Figure S 12: **Excess photo-assisted shot noise of Levitons**

An increase in temperature has a smoothing effect on the curves.

*4. Simulating multiple-harmonic photon-assisted shot noise*

For a single sine excitation sent to the sample, the complex probability amplitudes $p_l$ to shift the Fermi sea by $lh\nu$ are $p_l = J_l(\alpha)$ where $\alpha = \frac{eV_{RF}}{h\nu}$. However, they differ when more than one sine signal is sent at the same time. Since the $p_l$ are the Fourier coefficients of $e^{i\varphi(t)}$ where $\varphi$ is the phase acquired by the electrons as they reach the valley splitter, adding other harmonics at different frequencies will create additional phases and thus a product of complex phase factors. The coefficients of this new Fourier transform will thus be the convolution of the coefficients of the Fourier transforms of the original waveforms, accounting for different periodicities. Algorithmically, these multiple convolutions may be computed one after the other by noting that the formula given

16

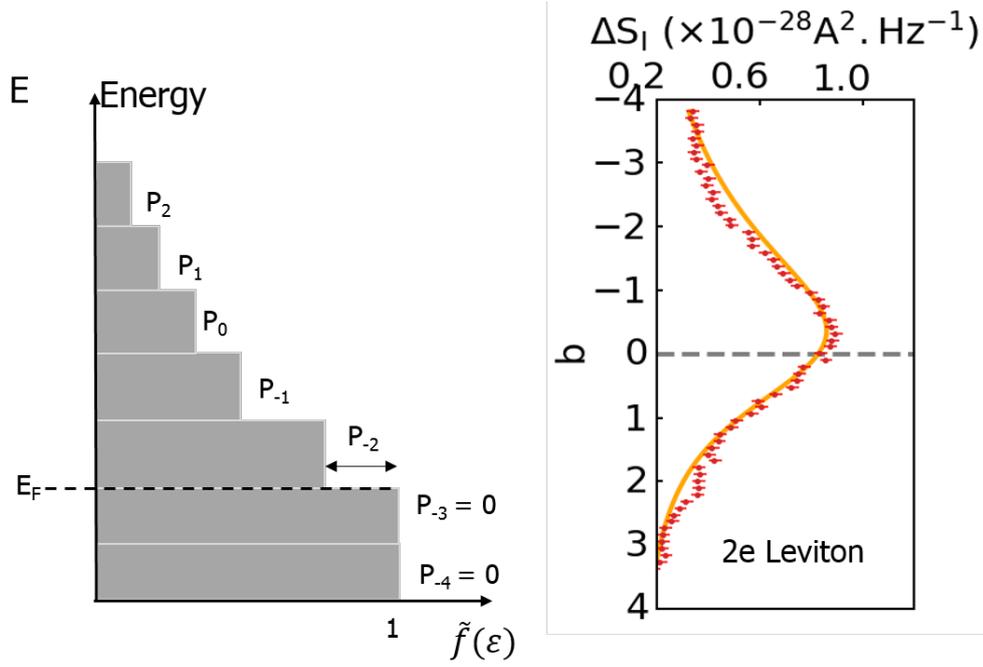

Figure S 13: **Measured shot noise of a 2e Leviton**

Distribution function for a 2e-Leviton and corresponding excess noise as a function of q for a 2e Leviton. The noise derivative is proportional to $f(\tilde{\epsilon})$ for $q < 0$ and to $-(1 - f(\tilde{\epsilon}))$ for $q > 0$.

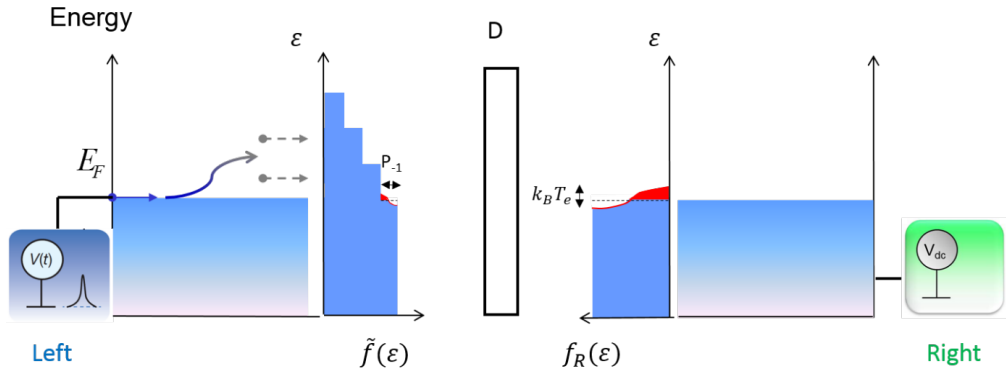

Figure S 14: **Energy spectroscopy**

Leviton pulse applied on the left ohmic contact modifies the Fermi distribution $\tilde{f}_L$. DC bias voltage $V_R$ on the right contact is swept to perform the energy spectroscopy.

in [47] for the $p_l$ amplitudes of a biharmonic signal easily generalizes to any two signals of respective frequency $\nu$ and $n \cdot \nu$ where $n$ is an integer : $p_l^{sum} = \sum_{k \in \mathbb{Z}} p_{l-n \cdot k} p'_k e^{-ik\varphi}$ with $p_l$ the coefficients of the signal of frequency $\nu$, $p'_l$ those of the signal of frequency $n \cdot \nu$ and $\varphi$ the phase offset of the harmonic with respect to the base signal. Thus, the action of summing harmonics in the time

17

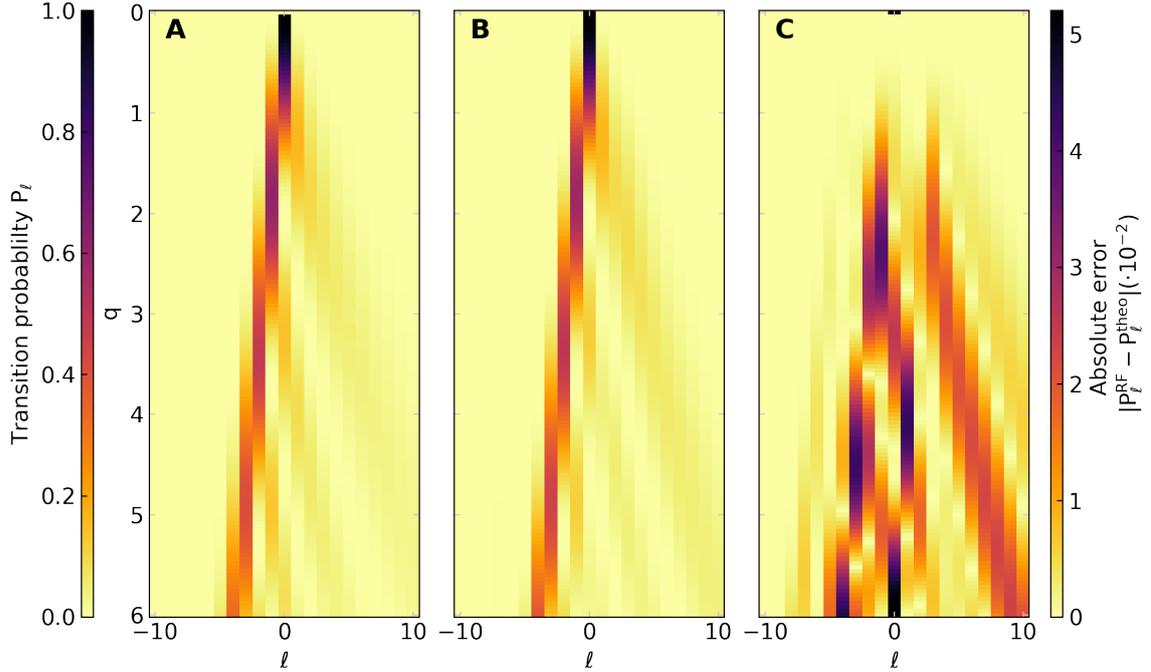

Figure S 15: **Comparison of transition probabilities** $P_l$

**A**: for an $\alpha = q$ Lorentzian. **B**: for its three-harmonic approximation. **A** and **B** share the same color scale which is on the left-hand side of the figure. **C**: absolute value of the difference between **A** and **B**, with its own color scale on the right-hand side of the figure. Apart from an artifact of the computation of **A** at $q = 0$, the maximal error due to the missing harmonics in the reconstruction has a value of about 0.052. Thus, at $\frac{W}{T} = 0.09$, three harmonics are sufficient to faithfully reconstruct the $P_l$ distribution of a train of Lorentzian pulses.

domain corresponds to performing a convolution in the domain of the $P_l$.

5. *Lorentzian pulses in a Mach-Zehnder interferometer: a simple approach*

Tuning both side gates to nonzero transmission, the excitation can be sent in the resulting Mach-Zehnder interferometer and its coherence properties can be studied. If the pulses sent in the interferometer keep their coherence throughout their whole path, the shot noise is essentially determined by the total number of electron-hole pairs created by the pulse multiplied by the factor $T_{MZI}$ (1-$T_{MZI}$) (a more detailed calculation is performed in "Floquet scattering formalism for graphene Mach Zehnder interferometry"). An example is shown at zero temperature in fig. S17 where the RF part of the pulse is that of a 1-electron Leviton. Fig. S18 illustrates the good accordance between our measurements and the theoretical predictions for shot noise of multiharmonic signals travelling through a MZI, here in the biharmonic case. It also shows that the shape of

18

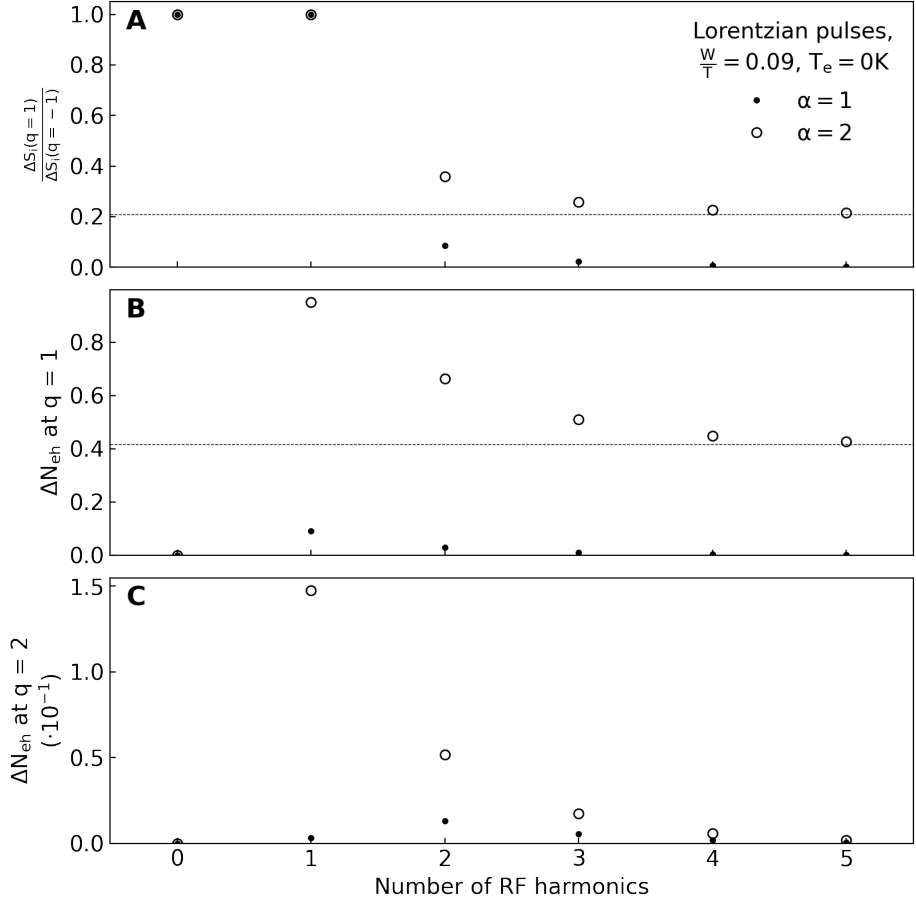

Figure S 16: **Features of the truncated Lorentzian pulse**

**A**: Zero-temperature asymmetry ratio $\mathcal{A} = \frac{S(q=1)}{S(q=-1)}$ of the excess photo-assisted shot noise. **B** and **C**: excess number of electron-hole pairs created (respectively) at $q = 1$ and $q = 2$ as the maximum order of truncation of the Fourier series of a 1-electron (resp. 2-electron) Lorentzian is varied. The two-electron Leviton's excess noise does not vanish at $q = 1$ and features worse asymmetry at that point, but adding more RF harmonics still converges towards the expected values for a 2-electron leviton (dashed horizontal lines). For the 1-electron leviton approximation, these metrics converge towards 0. At $q = 2$, both the 1-electron and 2-electron levitons converge towards zero excess noise.

pulses will remain the same while propagating through the interferometer, as no additional phase shift is introduced between the harmonics.

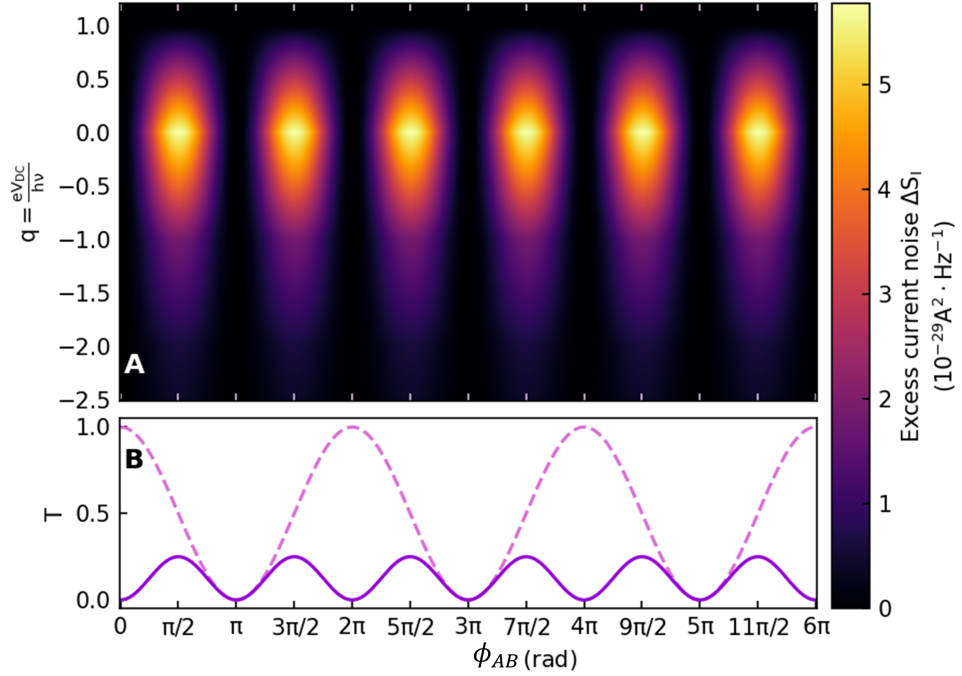

Figure S 17: **Simulation of the Aharonov-Bohm oscillations of the Mach-Zehnder interferometer when the Leviton wavepacket remains coherent throughout its travel in the arms**

**A**: Voltage-dependence of the excess photon-assisted shot noise at zero temperature for a one-electron Lorentzian excitation as the Aharonov-Bohm phase is swept. **B**: Corresponding evolution of the MZI transmission $T_{MZI}$ (dashed line) and the $T_{MZI}(1 - T_{MZI})$ prefactor governing the excess shot noise amplitude (full line) for both side gates tuned to 0.5.

### D. Number of electron-hole pairs and consequence on the electronic coherence

#### 1. Number of electron-hole pairs at finite temperature

We measure the excess noise $\Delta S_I$ obtained by subtracting the noise with $V_{ac}$ "off" from the one with $V_{ac}$ "on". At zero temperature, $\Delta S_I$ measures the excess particle number $\Delta N_{eh}$ which is defined as $\Delta N_{eh} = \Delta S_I / S_I^0 = N_e + N_h - n = \sum_l |l+n| P_l - |n|$, where $S_I^0 = 2\frac{2e^2}{h} D(1-D) h\nu$ is the typical scale of the photo-assisted shot noise. This quantity becomes null for Levitons with integer charge n. Note that for non-integer pulses, a finite number of electron-hole excitations is necessary to build the electron state even for Lorentzian pulses. At finite temperature, $\Delta N_{eh}$ becomes $\Delta N_{eh} = \sum_{l=-\infty}^{+\infty} (l+n) \times P_l \times \coth(\frac{l+q}{2\theta_e}) - q \times \coth(\frac{l+q}{2\theta_e})$ with $\theta_e = k_B T_e / h\nu \ll 1$. Thermal excitations now add an extra contribution to the excess noise. In the following, we will show that

20

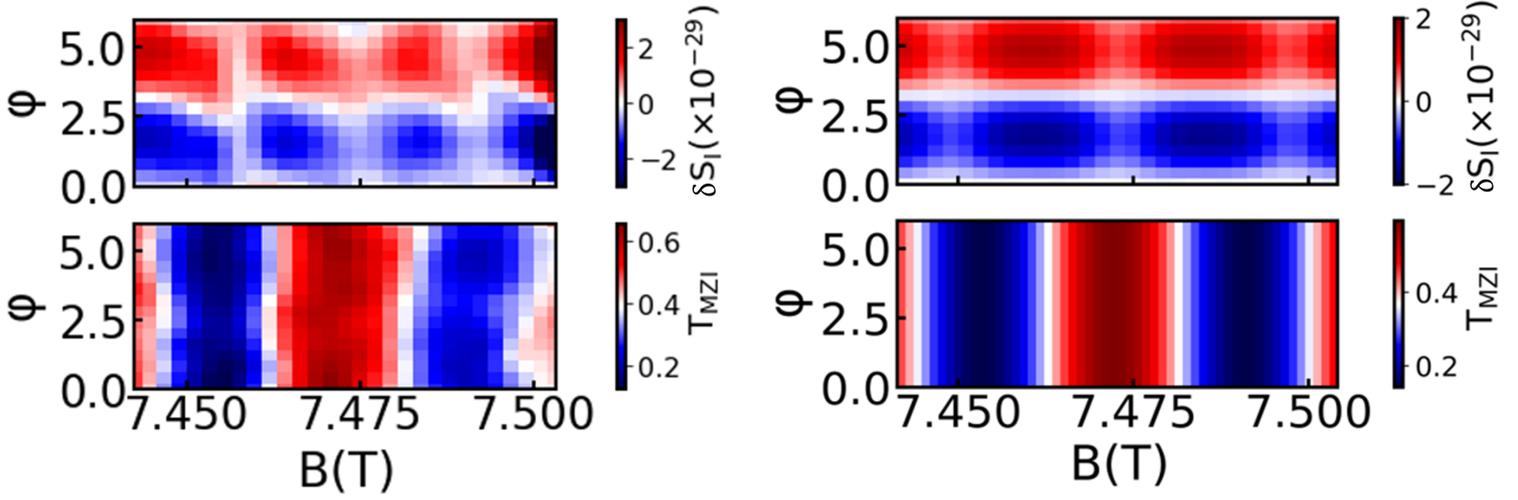

Figure S 18: **Measured and theoretical shot noise of a biharmonic excitation travelling through the MZI, as a function of the relative phase of the two harmonics and the magnetic field**

**Left**: A biharmonic excitation $V_{ac,bi}(t) = \alpha_1 cos(2\pi * \nu t) + \alpha_2 cos(2\pi * 2\nu t + \varphi)$ propagates in the MZI. $\delta S_I = S_I(q=1, \alpha_1, \alpha_2, \varphi) - S_I(q=-1, \alpha_1, \alpha_2, \varphi)$ is measured as a function of $\varphi$ and the magnetic field. The behavior remains the same on a 60mT magnetic field range, confirming that no dephasing between harmonics is introduced in the MZI. **Right**: theoretical prediction.

this extra contribution is $\sim 2\theta_e P_{-n}$.

The energy distribution of the right reservoir, $f_R(\epsilon)$, is given by a Fermi distribution at temperature $T_e$ (see also S14). The energy distribution $\tilde{f}_L(\epsilon)$ of electrons driven by the Lorentzian pulse at the left reservoir is $\tilde{f}_L(\epsilon) = \sum_l P_l f(\epsilon - l - n)$. The contribution to the noise of right and left electrons at energy $\epsilon$ is proportional to $s(\epsilon) = f_R(\epsilon)(1-\tilde{f}_L(\epsilon)) + \tilde{f}_L\epsilon(1-f_R(\epsilon))$. The integrated noise over the energy range $\Delta\epsilon$ with $\theta_e \ll \Delta\epsilon \ll 1$ gives $S_{\Delta\epsilon} = \Delta\epsilon \sum_{l>-n} P_l + 2\theta_e P_{-n}$ (for the n-particle Levitons, $\sum_{l=-n}^{\infty} P_l = 1$). The first term $\Delta\epsilon \sum_{l>-n} P_l$ corresponds to the zero temperature contribution. We finally obtain $\Delta N_{eh}(T_e) \sim \Delta N_{eh}(0) + 2\theta_e P_{-n}$.

*2. Levitons and decoherence*

In ref. [45] (more precisely in Fig. 7), the excess particle number is plotted as a function of the charge per period q for a Lorentzian pulse and a rectangular pulse at zero temperature. For rectangular pulse, $\Delta N_{eh}$ dramatically depends on the pulse shape and reaching $\sim 0.2$ for integer charge per period (1 or 2). For n=1 charge per period, this leads to the emission of an uncontrolled electron-hole pair for every 5 on-demand electrons.

In the following, we discuss the effect of this uncontrolled electron-hole pair on decoherence.

Single-electron interference:

In Ref. [48], the various sources of decoherence in an electronic MZI are studied. In MZI defined by edge states, decoherence primarily stems from intra-channel Coulomb interactions within a single edge channel or inter-channel Coulomb interactions between adjacent edge channels. We have shown that the latter one is not the dominant mechanism compared to the intra-channel Coulomb interactions. This is mainly due to the sample geometry, where the vertical distance (30 - 50nm) between the gates and the graphene layer is shorter than the spacing (50 - 60nm) between two adjacent edge channels, leading to screening of the inter-channel interactions. The intra-channel interaction Hamiltonian is given by $H_{\alpha,intra} = \frac{gv\hbar}{2L} (\int_0^L \rho_\alpha(x)dx - N_{g,\alpha})^2$, where $\alpha$ stands for the left ($\alpha$=l) or right ($\alpha$=r) arm of the MZI, g is the dimensionless interaction strength, v is the drift velocity, L is the arm length, and $N_{g,\alpha}$ is an electron number tuned by the gate voltage. Although the extension of this model to the finite frequency regime is missing, the intra-channel interaction Hamiltonian, and consequently the decoherence, should not be sensitive to the presence of neutral electron-hole excitations, as it does not affect the electron number.

Two-electron interference:

For two-electron operations, the presence of electron-hole pairs has a direct impact on the decoherence. For the Hong-Ou-Mandel experiment, shot noise can be expressed as $\frac{S_I}{S_I^0} = 2\sum_{k=1}^{\infty} |k||\Pi_k|^2$ with $\Pi_k = e^{-i2\pi\nu k\tau/2}(\sum_l p_l p_{l-k}^* e^{i2\pi\nu l\tau})$, $\nu$ the repetition frequency of the pulse and $\tau$ the time delay between the two pulses [45]. For Leviton, it can be shown that $\Pi_0 = \frac{1-(2-e^{i2\pi\nu\tau})\beta^2}{1-\beta^2 e^{i2\pi\nu\tau}}$ and $\Pi_{l\geq 1} = \beta^l(1-\beta^2)e^{-i2\pi l\nu\tau/2}\frac{e^{i2\pi\nu\tau}-1}{1-\beta^2 e^{i2\pi\nu\tau}}$ This finally gives $\frac{S_I}{S_I^0} = (2\sum_{l=1}^{\infty} l\beta^{2l}) |\frac{e^{i2\pi\nu\tau}-1}{1-\beta^2 e^{i2\pi\nu\tau}}|^2 (1-\beta^2)^2$ with $\beta$=exp(-2$\pi\eta$) and $\eta$=W/T (W is the width of the periodic pulse and T=1/$\nu$ is its period). Tem-

perature dependence can be incorporated by replacing l with l×coth($\frac{lh\nu}{2k_BT_e}$)-2k$_B$T$_e$/h$\nu$. As a result, the temperature dependence of the HOM shot noise decouples from the $\tau$ dependence and one can write S$_I$(T,$\tau$)=F(T) S$_I$ (0,$\tau$). Therefore, the temperature only reduces the noise but does not affect its shape, which is a unique property of the Lorentzian excitation. For other pulse shape, spurious electron-hole excitations will collide with thermal excitations, leading to an extra contribution to the noise. This decoherence process modifies the HOM dip.

3. *Relevance of the electronic interactions:*

In the quantum Hall regime at filling factor 2, two types of interactions are dominant.
- Inter-channel Coulomb interactions between adjacent edge channels This results in new propagating eigenmodes with different velocities, namely a fast charge mode and a slow neutral mode. This effect is a major source of decoherence and relaxation in conventional semiconductors as GaAs.
- Intra-channel Coulomb interactions within a single edge channel This can also be a source of decoherence. Physically, when an electron enters an interferometer arm, it causes charge density fluctuations in the arm, which provide 'which-path' information through the capacitive interaction, resulting in suppression of coherence.

In a recent study [48], it has been shown that intra-channel Coulomb interactions dominate over inter-channel interactions. To understand this, one has to consider the sample geometry in detail. The vertical distance (30 - 50 nm) (Fig. S1d) between the gates and the graphene layer is shorter than the spacing (50 - 60 nm) between two adjacent edge channels, leading to screening of the inter-channel interactions. Note that the spacing of 110 nm between the two arms is estimated assuming a symmetric PN junction from the Aharonov-Bohm period. This is in sharp contrast with the GaAs edge channels, where the distance between the edge channels and their distance to the top gate are both typically around 90 - 100 nm. The effect of intra-channel interaction appears in the visibility of Aharonov-Bohm oscillations as a function of the applied DC bias V$_{dc}$. By increasing V$_{dc}$, the charge density in each arm increases, leading to enhanced decoherence and resulting in a lobe pattern of the visibility as a function of V$_{dc}$ (Fig. 4 in Ref. [48]). The dip in visibility occurs around 220 $\mu$V, corresponding to a period between two consecutive electrons $\tau_{dc} \sim$ h/eV=19 ps. For one and two electron Levitons, we did not observe any deviation from the non-interacting theory. For a frequency $\nu$=3.5 GHz pulse, the period between two consecutive pulses is $\tau_{RF} \sim 1/\nu$=285 ps. $\tau_{RF} \ll \tau_{dc}$ combined with the fact that W/T=0.09 (W the width of the pulse and T=1/$\nu$) may explain why we did not observe any clear signature of the intra-channel

23

interactions with Levitons.

## II. THEORETICAL MODEL

### A. Floquet scattering formalism for graphene Mach-Zehnder interferometry

We derive the transmission and noise of the graphene Mach-Zehnder interferometer (MZI), based on the Floquet scattering formalism and taking the valley isospins into account.

#### 1. Setup

We consider a MZI on a graphene pn-junction. The valley isospin states on each interferometer arm and graphene edge are denoted in Fig. S19. The valley isospin vectors depend on the edge type. Generally, except a pristine zigzag edge (with a sublattice symmetry), chiral edge states coming from the left and the right to the MZI have opposite valley isospins, denoted by $\vec{w}_{\text{in}}$ and $-\vec{w}_{\text{in}}$ [49]; the chiral edge states have opposite valley isospins also for zigzag edges but with a broken sublattice symmetry (which is realistic) [50]. Similarly, the left and the right chiral edges outgoing from the MZI have another opposite valley isospins $\vec{w}_{\text{out}}$ and $-\vec{w}_{\text{out}}$, respectively. The MZI arms have opposite valley isospins $\vec{w}$ and $-\vec{w}$. We omit the (real) spin degree of freedom, which is preserved in the MZI and provides only degenerate states.

We introduce scattering matrices taking the valley isospin into account. Reflecting the configuration of the valley isospins shown in Fig. S19, the scattering matrices $S_1$ and $S_2$ at the top and

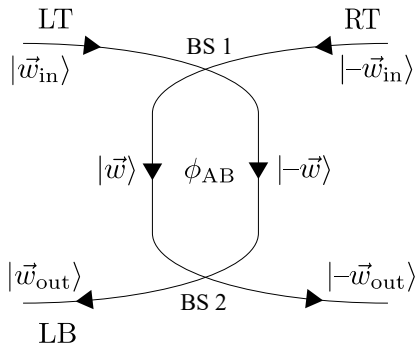

Figure S 19: Mach-Zehnder interferometer formed in a graphene pn junction. $|\vec{w}\rangle$'s denote the valley isospin states on each path.

24

the bottom beam splitters (BS1 and BS2) can be written as

$$S_1 = \begin{pmatrix} r_1 U_{-\vec{w}_{\text{in}},-\vec{w}} & t'_1 U_{-\vec{w}_{\text{in}},\vec{w}} \\ t_1 U_{\vec{w}_{\text{in}},-\vec{w}} & r'_1 U_{\vec{w}_{\text{in}},\vec{w}} \end{pmatrix}, \quad S_2 = \begin{pmatrix} r_2 U_{-\vec{w},-\vec{w}_{\text{out}}} & t'_2 U_{-\vec{w},\vec{w}_{\text{out}}} \\ t_2 U_{\vec{w},-\vec{w}_{\text{out}}} & r'_2 U_{\vec{w},\vec{w}_{\text{out}}} \end{pmatrix}. \tag{2}$$

At the top (resp. bottom) beam splitter, the electrons are transmitted with the probability $|t_1|^2$ ($|t_2|^2$) and reflected with the probability $|r_1|^2 = 1 - |t_1|^2$ ($|r_2|^2 = 1 - |t_2|^2$). The amplitudes $t_j, r_j, t'_j, r'_j$ must satisfy the unitarity condition $S_j^\dagger S_j = S_j S_j^\dagger = I$ and are assumed to be independent of electron energies. The $2 \times 2$ matrices $U_{\vec{a},\vec{b}}$ are the unitary transformations that rotate the valley isospin from $\vec{a}$ to $\vec{b}$. For example, $U_{\vec{w}_{\text{in}},\vec{w}}$ transforms $|\vec{w}_{\text{in}}\rangle$ to $|\vec{w}\rangle$ up to some $U(1)$ phase, which reflects the $4\pi$-periodic spin-1/2 structure. For a concreteness, we denote the phase factors that follow from the rotation due to the spin structure as follows:

$$U_{-\vec{w}_{\text{in}},-\vec{w}}|-\vec{w}_{\text{in}}\rangle = e^{i\Theta_{-\vec{w}_{\text{in}},-\vec{w}}}|-\vec{w}\rangle, \quad U_{-\vec{w}_{\text{in}},\vec{w}}|-\vec{w}_{\text{in}}\rangle = e^{i\Theta_{-\vec{w}_{\text{in}},\vec{w}}}|\vec{w}\rangle, \tag{3}$$

$$U_{\vec{w}_{\text{in}},-\vec{w}}|\vec{w}_{\text{in}}\rangle = e^{i\Theta_{\vec{w}_{\text{in}},-\vec{w}}}|-\vec{w}\rangle, \quad U_{\vec{w}_{\text{in}},\vec{w}}|\vec{w}_{\text{in}}\rangle = e^{i\Theta_{\vec{w}_{\text{in}},\vec{w}}}|\vec{w}\rangle, \tag{4}$$

$$U_{-\vec{w},-\vec{w}_{\text{out}}}|-\vec{w}\rangle = e^{i\Theta_{-\vec{w},-\vec{w}_{\text{out}}}}|-\vec{w}_{\text{out}}\rangle, \quad U_{-\vec{w},\vec{w}_{\text{out}}}|-\vec{w}\rangle = e^{i\Theta_{-\vec{w},\vec{w}_{\text{out}}}}|\vec{w}_{\text{out}}\rangle, \tag{5}$$

$$U_{\vec{w},-\vec{w}_{\text{out}}}|\vec{w}\rangle = e^{i\Theta_{\vec{w},-\vec{w}_{\text{out}}}}|-\vec{w}_{\text{out}}\rangle, \quad U_{\vec{w},\vec{w}_{\text{out}}}|\vec{w}\rangle = e^{i\Theta_{\vec{w},\vec{w}_{\text{out}}}}|\vec{w}_{\text{out}}\rangle. \tag{6}$$

The Aharonov-Bohm phase due to the magnetic field can be described by

$$S_\phi = \begin{pmatrix} e^{i\phi_{\text{AB}}} I & 0 \\ 0 & I \end{pmatrix}. \tag{7}$$

The total scattering matrix of the MZI for each energy mode then becomes $S_2 S_\phi S_1$. Here we assume no length difference between the two arms enclosing the magnetic flux.

We now derive a scattering formalism in our experimental situation where a periodic voltage $V(t) = V_{\text{dc}} + V_{\text{ac}}(t)$ with frequency $\nu$ is applied at either the left top (LT) or the right top (RT) contact. The pulse is parametrized to carry a charge $q$ per period $\mathcal{T} = 1/\nu$, i.e., $eV_{\text{dc}} = qh\nu$ and $\int_0^\mathcal{T} dt V_{\text{ac}}(t) = 0$. For the purpose, we combine the aforementioned scattering matrix with the Floquet amplitudes induced by the AC voltage $V_{\text{ac}}(t)$. The amplitude that an energy mode $E$ is scattered to another energy mode $E_n = E + nh\nu$ is given by

$$\mathcal{S}_n = \int_0^\mathcal{T} \frac{dt}{\mathcal{T}} e^{2\pi i n \nu t} e^{-i\varphi(t)} \tag{8}$$

where $\varphi(t) = \frac{e}{\hbar} \int_0^t dt' V_{\text{ac}}(t')$. Then, the Floquet scattering matrices that relate incoming energy modes $E_m$ at the left top contact (LT) to outgoing energy modes $E_n$ toward the left bottom (LB)

contact become

$$\mathcal{F}^{\mathrm{L}}_{\mathrm{LT}\to\mathrm{LB}}(E_n, E_m) = \begin{pmatrix} 0 & \langle \vec{w}_{\mathrm{out}}| \end{pmatrix} \mathcal{S}^{\mathrm{L}}_{\mathrm{LT},n-m} S_2 S_\phi S_1 \begin{pmatrix} 0 \\ |\vec{w}_{\mathrm{in}}\rangle \end{pmatrix}$$

$$= \mathcal{S}^{\mathrm{L}}_{\mathrm{LT},n-m} \left[ t'_1 t_2 e^{i(\phi_{\mathrm{AB}} + \Theta_{\vec{w}_{\mathrm{in}},-\vec{w}} + \Theta_{-\vec{w},\vec{w}_{\mathrm{out}}})} + r'_1 r'_2 e^{i(\Theta_{\vec{w}_{\mathrm{in}},\vec{w}} + \Theta_{\vec{w},\vec{w}_{\mathrm{out}}})} \right], \quad (9)$$

$$\mathcal{F}^{\mathrm{L}}_{\mathrm{RT}\to\mathrm{LB}}(E_n, E_m) = \begin{pmatrix} 0 & \langle \vec{w}_{\mathrm{out}}| \end{pmatrix} \mathcal{S}^{\mathrm{L}}_{\mathrm{RT},n-m} S_2 S_\phi S_1 \begin{pmatrix} |-\vec{w}_{\mathrm{in}}\rangle \\ 0 \end{pmatrix}$$

$$= \delta_{n,m} \left[ r_1 t_2 e^{i(\phi_{\mathrm{AB}} + \Theta_{-\vec{w}_{\mathrm{in}},-\vec{w}} + \Theta_{-\vec{w},\vec{w}_{\mathrm{out}}})} + t_1 r'_2 e^{i(\Theta_{-\vec{w}_{\mathrm{in}},\vec{w}} + \Theta_{\vec{w},\vec{w}_{\mathrm{out}}})} \right] \quad (10)$$

where the supersript L indicates the left charge injection. The Floquet amplitude $\mathcal{S}^{\mathrm{L}}_{\mathrm{LT},n-m}$ corresponds to the AC voltage $V_{\mathrm{ac}}(t)$ applied at the left top (LT) contact, while $\mathcal{S}^{\mathrm{L}}_{\mathrm{RT},n-m} = \delta_{n,m}$ corresponds to no AC voltage at the right top (RT). We have $\mathcal{S}^{\mathrm{L}}_{\mathrm{LT},n-m} = \mathcal{S}^{\mathrm{R}}_{\mathrm{RT},n-m} \equiv \mathcal{S}_{n-m}$ since we inject the same voltage pulse for the left and the right injection. Similarly, we write the following Floquet scattering matrices relating incoming energy modes $E_m$ at the right top contact (RT) to outgoing energy modes $E_n$ toward the left bottom (LB):

$$\mathcal{F}^{\mathrm{R}}_{\mathrm{LT}\to\mathrm{LB}}(E_n, E_m) = \begin{pmatrix} 0 & \langle \vec{w}_{\mathrm{out}}| \end{pmatrix} \mathcal{S}^{\mathrm{R}}_{\mathrm{LT},n-m} S_2 S_\phi S_1 \begin{pmatrix} 0 \\ |\vec{w}_{\mathrm{in}}\rangle \end{pmatrix}$$

$$= \delta_{n,m} \left[ t'_1 t_2 e^{i(\phi_{\mathrm{AB}} + \Theta_{\vec{w}_{\mathrm{in}},-\vec{w}} + \Theta_{-\vec{w},\vec{w}_{\mathrm{out}}})} + r'_1 r'_2 e^{i(\Theta_{\vec{w}_{\mathrm{in}},\vec{w}} + \Theta_{\vec{w},\vec{w}_{\mathrm{out}}})} \right] \quad (11)$$

$$\mathcal{F}^{\mathrm{R}}_{\mathrm{RT}\to\mathrm{LB}}(E_n, E_m) = \begin{pmatrix} 0 & \langle \vec{w}_{\mathrm{out}}| \end{pmatrix} \mathcal{S}^{\mathrm{R}}_{\mathrm{RT},n-m} S_2 S_\phi S_1 \begin{pmatrix} |-\vec{w}_{\mathrm{in}}\rangle \\ 0 \end{pmatrix}$$

$$= \mathcal{S}^{\mathrm{R}}_{\mathrm{RT},n-m} \left[ r_1 t_2 e^{i(\phi_{\mathrm{AB}} + \Theta_{-\vec{w}_{\mathrm{in}},-\vec{w}} + \Theta_{-\vec{w},\vec{w}_{\mathrm{out}}})} + t_1 r'_2 e^{i(\Theta_{-\vec{w}_{\mathrm{in}},\vec{w}} + \Theta_{\vec{w},\vec{w}_{\mathrm{out}}})} \right]. \quad (12)$$

2. Current

When the AC voltage is applied at the left top contact, the transmission $T^L$ measured at the left bottom contact is given by

$$T^{\mathrm{L}} = \frac{e}{h} \int dE \left[ \sum_n \sum_{\alpha=\mathrm{LT,RT}} |\mathcal{F}^{\mathrm{L}}_{\alpha\to\mathrm{LB}}(E, E_n)|^2 f^{\mathrm{L}}_\alpha(E_n) - f_{\mathrm{LB}}(E) \right] \quad (13)$$

where $f^{\mathrm{L}}_{\mathrm{LT}}(E_n) = f(E_n - eV_{\mathrm{dc}})$, $f^{\mathrm{L}}_{\mathrm{RT}}(E_n) = f(E_n)$, $f_{\mathrm{LB}}(E) = f(E)$, and $f(E)$ is the Fermi-Dirac distribution. Similarly, when the AC voltage is applied at the right top contact, the transmission $T^R$ measured at the left bottom contact is given by

$$T^{\mathrm{R}} = \frac{e}{h} \int dE \left[ \sum_n \sum_{\alpha=\mathrm{LT,RT}} |\mathcal{F}^{\mathrm{R}}_{\alpha\to\mathrm{LB}}(E, E_n)|^2 f^{\mathrm{R}}_\alpha(E_n) - f_{\mathrm{LB}}(E) \right] \quad (14)$$

where $f_{\text{RT}}^{\text{R}}(E_n) = f(E_n - eV_{\text{dc}})$ and $f_{\text{LT}}^{\text{R}}(E_n) = f(E_n)$. Using the relation

$$\frac{e}{h}\int dE \left(\sum_n |S_n|^2 f(E_{-n} - eV_{\text{dc}}) - f(E)\right) = \frac{qe}{\mathcal{T}}, \tag{15}$$

the transmissions are found as

$$T^{\text{L}} = \frac{qe}{\mathcal{T}}\left[|r_1'|^2|r_2'|^2 + |t_1'|^2|t_2|^2 + 2|r_1'r_2't_1't_2|\cos\bigl(\phi_{\text{AB}} + \arg(t_1't_2 r_1'^* r_2'^*) + \gamma_{\text{L}}\bigr)\right], \tag{16}$$

$$T^{\text{R}} = \frac{qe}{\mathcal{T}}\left[|t_1|^2|r_2'|^2 + |r_1|^2|t_2|^2 + 2|t_1 r_2' r_1 t_2|\cos\bigl(\phi_{\text{AB}} + \arg(r_1 t_2 t_1^* r_2'^*) + \gamma_{\text{R}}\bigr)\right]. \tag{17}$$

Here the phases due to valley isospin rotation

$$\gamma_{\text{L}} = \Theta_{\vec{w}_{\text{in}},-\vec{w}} + \Theta_{-\vec{w},\vec{w}_{\text{out}}} - \Theta_{\vec{w},\vec{w}_{\text{out}}} - \Theta_{\vec{w}_{\text{in}},\vec{w}} \tag{18}$$

$$\gamma_{\text{R}} = \Theta_{-\vec{w}_{\text{in}},-\vec{w}} + \Theta_{-\vec{w},\vec{w}_{\text{out}}} - \Theta_{\vec{w},\vec{w}_{\text{out}}} - \Theta_{-\vec{w}_{\text{in}},\vec{w}} \tag{19}$$

are the Pancharatnam-Berry phases corresponding to the blue and red trajectories on the Bloch sphere shown in Fig. S20. These phases are identical to the minus half of the solid angles of the closed loop corresponding to the valley isospin rotation [51]. Notice that the difference between $\gamma_{\text{L}}$ and $\gamma_{\text{R}}$ equals $\pi$,

$$\gamma_{\text{L}} - \gamma_{\text{R}} = \pi \quad (\text{mod } 2\pi). \tag{20}$$

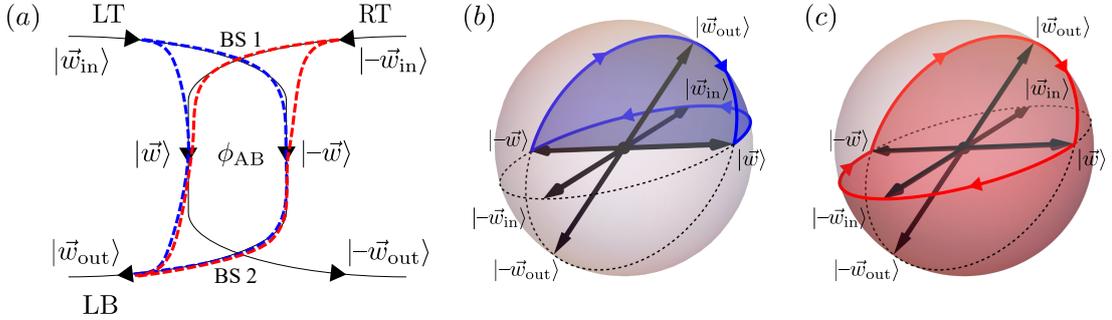

Figure S 20: **A**: interference of an electron injected from the left top contact LT involves the blue closed loop, while that of an electron from the right top contact RT involves the red closed loop. **B**: along the blue loop for the left injection, the valley isospin sequentially rotates as $\vec{\omega}_{\text{in}} \to -\vec{\omega} \to \vec{\omega}_{\text{out}} \to \vec{\omega} \to \vec{\omega}_{\text{in}}$, resulting in a solid angle $-2\gamma_{\text{L}}$ on the Bloch sphere. The interference by the blue loop has the phase shift of $\gamma_{\text{L}}$. **C**: Along the red loop for the right injection, the valley isospin sequentially rotates as $-\vec{\omega}_{\text{in}} \to -\vec{\omega} \to \vec{\omega}_{\text{out}} \to \vec{\omega} \to -\vec{\omega}_{\text{in}}$, resulting in a solid angle $-2\gamma_{\text{R}}$. The interference by the red loop has the phase shift of $\gamma_{\text{R}}$. The difference between the two solid angles $-2\gamma_{\text{L}}$ and $-2\gamma_{\text{R}}$ equals $2\pi$, the solid angle of a hemisphere.

*3. Noise*

For an MZI of equal arm lengths, the main contribution of the shot noise is the transport noise $\mathcal{P}$, which originates from the two different input contact [52]. For the injection from the left top contact, the zero-frequency auto-correlation measured at the left bottom contact is given by [53]

$$\mathcal{P}^{\mathrm{L}} = \frac{e^2}{h} \int dE \sum_{\{\alpha,\beta\}=\{\mathrm{LT,RT}\}} \sum_{n,m,p} \frac{[f_\alpha^{\mathrm{L}}(E_n) - f_\beta^{\mathrm{L}}(E_m)]^2}{2}$$
$$\times [\mathcal{F}_{\alpha\to\mathrm{LB}}^{\mathrm{L}}(E,E_n)]^* \mathcal{F}_{\beta\to\mathrm{LB}}^{\mathrm{L}}(E,E_m) [\mathcal{F}_{\beta\to\mathrm{LB}}^{\mathrm{L}}(E_p,E_m)]^* \mathcal{F}_{\alpha\to\mathrm{LB}}^{\mathrm{L}}(E_p,E_n) \quad (21)$$

in the Floquet scattering formalism. Similarly, for the injection from the right top contact, the zero-frequency auto-correlation measured at the left bottom contact is given by

$$\mathcal{P}^{\mathrm{R}} = \frac{e^2}{h} \int dE \sum_{\{\alpha,\beta\}=\{\mathrm{LT,RT}\}} \sum_{n,m,p} \frac{[f_\alpha^{\mathrm{R}}(E_n) - f_\beta^{\mathrm{R}}(E_m)]^2}{2}$$
$$\times [\mathcal{F}_{\alpha\to\mathrm{LB}}^{\mathrm{R}}(E,E_n)]^* \mathcal{F}_{\beta\to\mathrm{LB}}^{\mathrm{R}}(E,E_m) [\mathcal{F}_{\beta\to\mathrm{LB}}^{\mathrm{R}}(E_p,E_m)]^* \mathcal{F}_{\alpha\to\mathrm{LB}}^{\mathrm{R}}(E_p,E_n). \quad (22)$$

Using the integration formula

$$\int dE [f(E_{-n} - eV_{\mathrm{dc}}) - f(E)]^2 = \bar{S}_0^{(n+q)}, \qquad \bar{S}_0^{(n)} = nh\nu \coth\left(\frac{nh\nu}{2k_B T}\right) - 2k_B T \quad (23)$$

with temperature $T$, we obtain

$$\mathcal{P}^L = \frac{e^2}{h} \sum_n |\mathcal{S}_n|^2 \bar{S}_0^{(n+q)} \left[c_0 - c_\phi \cos\bigl(\phi_{\mathrm{AB}} + \arg(t_1' t_2 r_1'^* r_2'^*) + \gamma_{\mathrm{L}}\bigr)\right.$$
$$\left. - c_{2\phi} \cos\bigl(2(\phi_{\mathrm{AB}} + \arg(t_1' t_2 r_1'^* r_2'^*) + \gamma_{\mathrm{L}})\bigr)\right], \quad (24)$$

$$\mathcal{P}^R = \frac{e^2}{h} \sum_n |\mathcal{S}_n|^2 \bar{S}_0^{(n+q)} \left[c_0 + c_\phi \cos\bigl(\phi_{\mathrm{AB}} + \arg(r_1 t_2 t_1^* r_2'^*) + \gamma_{\mathrm{R}}\bigr)\right.$$
$$\left. - c_{2\phi} \cos\bigl(2(\phi_{\mathrm{AB}} + \arg(r_1 t_2 t_1^* r_2'^*) + \gamma_{\mathrm{R}})\bigr)\right] \quad (25)$$

where

$$c_0 = |t_1|^2 |r_1|^2 + |t_2|^2 |r_2|^2 - 6|t_1|^2 |r_1|^2 |t_2|^2 |r_2|^2 \quad (26)$$
$$c_\phi = 2(|t_1|^2 - |r_1|^2)(|t_2|^2 - |r_2|^2)|r_1 r_2 t_1 t_2| \quad (27)$$
$$c_{2\phi} = 2|t_1|^2 |r_1|^2 |t_2|^2 |r_2|^2. \quad (28)$$

These results agree with the cross-correlation formula by Hofer and Flindt [52] (except the valley-induced Pancharatnam-Berry phase $\gamma_{\mathrm{R/L}}$).

28

### 4. Asymmetric MZI

For an asymmetric MZI or $\Delta_L \neq 0$ with $\Delta_L$ the length difference between the two arms, the transport noise has to be modified and an additional contribution to the noise called interference noise has to be taken into account [52]. The transport noise becomes:

$$\mathcal{P}^{R,as} = \frac{e^2}{h} \sum_n |\mathcal{S}_n|^2 \left[ c_0 \bar{S}_0^{(n+q)} + c_\phi \bar{S}_\phi^{(n+q)} \cos\left(\phi_{\text{AB}} + \arg(r_1 t_2 t_1^* r_2'^*) + \gamma_{\text{R}} + \pi(n+q)\nu\tau\right) \right.$$
$$\left. - c_{2\phi} \bar{S}_{2\phi}^{(n+q)} \cos\left(2(\phi_{\text{AB}} + \arg(r_1 t_2 t_1^* r_2'^*) + \gamma_{\text{R}} + 2\pi(n+q)\nu\tau)\right) \right], \tag{29}$$

$$\tag{30}$$

where $\tau = \Delta_L / v_D$, $v_D$ the drift velocity and:

$$\bar{S}_{j\phi}^{(n)} = \frac{2\pi k_B T}{\sinh(j\pi k_B T \tau/\hbar)} \left[ \coth\left(\frac{nh\nu}{2k_B T}\right) \sin(jn\pi\nu\tau) - \frac{jk_B T \tau}{\hbar} \cos(jn\pi\nu\tau) \right] \tag{31}$$

The interference noise writes [52]:

$$\mathcal{P}^{int} = \frac{e^2}{h} c_{2\phi} \sum_n \left[ \bar{S}_0^{(n)} |\mathcal{K}_n|^2 + \bar{S}_{2\phi}^{(n)} \operatorname{Re}\left\{ e^{i(2\phi_{\text{AB}} + 2(n+q)2\pi\nu\tau)} \mathcal{K}_n \mathcal{K}_{-n} \right\} \right]. \tag{32}$$

$$\tag{33}$$

. where $\mathcal{K}_n = \frac{1}{\sqrt{2}} \sum_m S_m^* S_{m+n} e^{im2\pi\nu\tau}$.

Finally, it can be shown that when the propagation velocity of outer (noted $v_o$) and inner (noted $v_i$) edge channel are different, electrons on the outer and inner edge channel have different dynamical phases $\varphi_o = 2\pi\nu\tau_o$ and $\varphi_i = 2\pi\nu\tau_i$ [54, 55]. In the presence of mixing points along the path, this will generate photo-assisted shot noise. Incoming electrons are noisy leading to an additional noise term:

$$\mathcal{P}^{inc} = \frac{e^2}{h} \left[ 2h\nu T_m R_m T_{MZI}^2 \right]. \tag{34}$$

$$\tag{35}$$

where ($T_m$, $R_m$) describes the edge tunnelling probabilities [54]. The total noise is the sum of these three terms: $\mathcal{P}^{tot} = \mathcal{P}^{R,as} + \mathcal{P}^{int} + \mathcal{P}^{inc}$

5. *Unitarity Condition*

Below we show that the relative Pancharatnam-Berry phase $\pi$ of the two loops (blue and red) shown in Fig. S20 is absorbed in the scattering matrix phases due to the unitarity condition. Imposing $S_j S_j^\dagger = S_j^\dagger S_j = I$, it must be

$$\begin{cases} r_1 t_1^* U_{-\vec{w}_{\text{in}},-\vec{w}} U_{\vec{w}_{\text{in}},-\vec{w}}^\dagger + t_1' r_1'^* U_{-\vec{w}_{\text{in}},\vec{w}} U_{\vec{w}_{\text{in}},\vec{w}}^\dagger = 0, \\ r_2 t_2^* U_{-\vec{w},-\vec{w}_{\text{out}}} U_{\vec{w},-\vec{w}_{\text{out}}}^\dagger + t_2' r_2'^* U_{-\vec{w},\vec{w}_{\text{out}}} U_{\vec{w},\vec{w}_{\text{out}}}^\dagger = 0. \end{cases} \quad (36)$$

Meanwhile, from the $4\pi$-periodicity of the spin structure, we have

$$U_{-\vec{w}_{\text{in}},-\vec{w}} U_{\vec{w}_{\text{in}},-\vec{w}}^\dagger U_{\vec{w}_{\text{in}},\vec{w}} U_{-\vec{w}_{\text{in}},\vec{w}}^\dagger = -I, \quad U_{-\vec{w},-\vec{w}_{\text{out}}} U_{\vec{w},-\vec{w}_{\text{out}}}^\dagger U_{\vec{w},\vec{w}_{\text{out}}} U_{-\vec{w},\vec{w}_{\text{out}}}^\dagger = -I, \quad (37)$$

which correspond to rotating a valley isospin by an angle $2\pi$. Combining Eq. (36) and Eq. (37), we obtain a constraint $r_j t_j^* - t_j' r_j'^* = 0$, or, equivalently,

$$\arg(t_j' t_j r_j^* r_j'^*) = 0 \quad (38)$$

for $j = 1, 2$, which is nothing but the relative difference between the scattering matrix phases in Eqs. (16) and (17). Combining Eq. (38) and $\gamma_\text{R} \equiv \gamma_\text{L} + \pi \pmod{2\pi}$ in Eq. (20), we find that the transmissions $T^\text{L}$ and $T^\text{R}$ for the left and the right injections, respectively, are out of phase. Meanwhile, we have $\mathcal{P}^L = \mathcal{P}^R$ for the noise, i.e., the noises from the left injection and the right injection are the same (or in phase).

Interestingly, the properties that the Aharonov-Bohm oscillations of $T^\text{L}$ and $T^\text{R}$ are out of phase and that $\mathcal{P}^L = \mathcal{P}^R$ (in phase) are also hold in the absence of the valley degree of freedom. It is because the unitarity of the scattering matrices results in $\arg(t_j' t_j r_j^* r_j'^*) = \pi$ instead of Eq. (38) in the absence of the valley isospins.

---



\end{document}